\documentclass[12pt]{iopart}
\usepackage{epsfig,graphicx}

\newcommand{\diag}{{\rm diag\,}}

\newcommand{\trg}{{\rm trg\,}}
\newcommand{\detg}{{\rm detg\,}}

\newcommand{\im}{{\rm Im\,}}

\begin{document}

\title[Arbitrary Rotation Invariant Random Matrix Ensembles 
         and Supersymmetry] {Arbitrary Rotation Invariant Random
         Matrix Ensembles and Supersymmetry}

\author{Thomas Guhr}
\address{Matematisk Fysik, LTH, Lunds Universitet,
                 Box 118, 22100 Lund, Sweden}

\begin{abstract}
We generalize the supersymmetry method in Random Matrix Theory to
arbitrary rotation invariant ensembles. Our exact approach further
extends a previous contribution in which we constructed a
supersymmetric representation for the class of norm--dependent Random
Matrix Ensembles.  Here, we derive a supersymmetric formulation under
very general circumstances. A projector is identified that provides
the mapping of the probability density from ordinary to
superspace. Furthermore, it is demonstrated that setting up the theory
in Fourier superspace has considerable advantages. General and exact
expressions for the correlation functions are given. We also show how
the use of hyperbolic symmetry can be circumvented in the present
context in which the non--linear $\sigma$ model is not used. We
construct exact supersymmetric integral representations of the
correlation functions for arbitrary positions of the imaginary
increments in the Green functions.
\end{abstract}




\section{Introduction}
\label{sec1}

The supersymmetry method is nowadays indispensable for the discussion
of various advanced topics in the theory of disordered
systems~\cite{EFE83,EFE97}, and it became equally important in
numerous random matrix approaches to complex systems in
general~\cite{VWZ,GMGW,Haake,TGenc}. Random Matrix Theory (RMT) as
originally formulated in ordinary space does not rely on Gaussian
probability densities. It is only important that the Random Matrix
Ensembles are invariant under basis rotations. Gaussian probability
densities are highly convenient in calculations, but other probability
densities are also possible, and some of those were already considered
in the early days of RMT~\cite{Mehta}. On the other hand, the
supersymmetric formulations were constructed for Gaussian probability
densities~\cite{EFE83,EFE97,ZIR04} by means of a Hubbard--Stratonovich
transformation. Thus, the question arises naturally whether the
Hubbard--Stratonovich transformation restricts the use of
supersymmetry to the Gaussian form of the probability densities.  We
address this problem in the present contribution. We will show that
the supersymmetry method is not at all restricted in this way, and we
will derive supersymmetric formulations of RMT for arbitrary rotation
invariant Random Matrix Ensembles.

We focus on conceptual and structural issues. In particular, we are
not aiming at asymptotic results in the inverse level number as
following from the supersymmetric non--linear $\sigma$
model~\cite{EFE83,EFE97,VWZ}.  This latter approach was used in
Ref.~\cite{HaWe} to show universality for infinite level--number in
the case of non--Gaussian probability densities.  Here, however, our
goal is different: we address the full problem to achieve exact,
i.e.~non--asymptotic results.  In a previous study~\cite{TGnorm}, we
presented supersymmetric representations for norm--dependent
ensembles, where the probability densities are functions of the traced
squared random matrices only. Although a series of interesting
insights are revealed already in this case, the derivation can be done
without actually employing deep features of supersymmetry.  This is
not so in the present contribution which aims at a general
construction. The methods needed are very different from the ones of
Ref.~\cite{TGnorm}. Here, we have to explore the algebraic structure
of superspace.

One can also motivate the present investigation from the viewpoint of
applications. We refer the interested reader to the
contribution~\cite{TGnorm} and the literature quoted therein. Our goal
to perform a conceptual study does not prevent us from giving general
expressions for the correlation functions, but we refrain from looking
too much into applications and defer this aspect to future work.

It will not be surprising for those who already have expertise in
supersymmetry that a generalization as outlined above requires an
analysis of convergence properties and thus leads inevitably to the
issue of what kind of symmetries the theory in superspace should have.
It was argued in Ref.~\cite{SW80} that hyperbolic symmetry,
i.e.~groups comprising compact and non--compact degrees of freedom,
are necessary if one is to set up a non--linear $\sigma$ model in
ordinary space. This line of reasoning carries over to
superspace~\cite{EFE83,EFE97}, see also the recent review in
Ref.~\cite{ZIR04}. We justify a procedure for how to avoid hyperbolic
symmetry in the framework of our supersymmetric models. The necessity
to introduce hyperbolic symmetry is exclusively rooted in the
non--linear $\sigma$ model, not in supersymmetry as such. If one aims
at exact, i.e.~non--asymptotic results, compact supergroups suffice.

For various reasons, including some related to convergence questions,
we find it advantageous to map the theory onto Fourier superspace.
Moreover, we restrict ourselves to unitary Random Matrix Ensembles
throughout the whole study.

The paper is organized as follows.  Having posed the problem in
Section~\ref{sec2}, we generalize the Hubbard--Stratonovich
transformation in Section~\ref{sec3}.  In Section~\ref{sec4}, we
derive the supersymmetric formulation in Fourier superspace. The
correlation functions are expressed as eigenvalue integrals in
Section~\ref{sec5}. Summary and conclusions are given in
Section~\ref{sec6}.

\section{Posing the Problem}
\label{sec2}

In Section~\ref{sec21}, the two relevant kinds of $k$--point
correlation functions are defined and the relation to the generating
functions is given. Thereby we also introduce our notation and
conventions. We clarify what we mean by arbitrary rotation invariant
ensembles in Section~\ref{sec22}.  In Section~\ref{sec23}, we show how
different types of correlation functions can be related to each other
by proper Fourier transforms.

\subsection{Correlation and Generating Functions}
\label{sec21}

The Random Matrix Ensemble builds upon $N\times N$ Hermitean matrices
$H$, having altogether $N^2$ independent matrix elements.  A
normalized probability density $P(H)$ assigns a statistical weight to
the elements of the matrices $H$. As the Hermitean matrices are
diagonalized by unitary matrices in ${\rm SU}(N)$, the probability
density $P(H)$ is said to define a Unitary Random Matrix Ensemble.  We
are interested in the $k$--point correlation functions
\begin{equation}
R_k(x_1,\ldots,x_k) = \int d[H] \, P(H) \prod_{p=1}^k \tr \delta\left(x_p-H\right) \ ,
\label{corrdefi}
\end{equation}
depending on the $k$ energies $x_1,\ldots,x_k$. The $\delta$ functions
are the imaginary parts of the matrix Green's functions, $\mp
i\pi\delta\left(x_p-H\right)=\im\left(x_p\pm i\varepsilon
-H\right)^{-1}$. Here, $i\varepsilon$ is an imaginary increment and
the limit $\varepsilon\to 0$ is suppressed in the notation. In the
supersymmetric construction to follow, it is convenient to consider
the more general correlation functions which also include the real
parts of the Green's functions.  They are, apart from an irrelevant
overall sign, given by
\begin{equation}
\widehat{R}_k(x_1,\ldots,x_k) = \frac{1}{\pi^k} \int d[H] \, P(H) 
                          \prod_{p=1}^k \tr \frac{1}{x_p-iL_p\varepsilon-H} \ .
\label{corrdef}
\end{equation}
One often wants to put the imaginary increments on different sides of
the real axis. The quantities $L_p$ which are either $+1$ or $-1$
determine the side of the real axis where the imaginary increment is
placed.  The correlation function can always be expressed as
derivatives of a generating function $Z_k(x+J)$ such that
\begin{equation}
\widehat{R}_k(x_1,\ldots,x_k) = 
                                      \frac{1}{(2\pi)^k} \frac{\partial^k}{\prod_{p=1}^k\partial J_p}
                                      Z_k(x+J) \Bigg|_{J_p=0}
\label{corr}
\end{equation}
where
\begin{equation}
Z_k(x+J) = \int d[H] \, P(H) \prod_{p=1}^k  
\frac{\det(H - x_p + iL_p\varepsilon - J_p)}{\det(H - x_p + iL_p\varepsilon + J_p)} \ .
\label{gen}
\end{equation}
We introduced source variables $J_p , \ p=1,\ldots,k$ as well as the
diagonal matrices $x=\diag(x_1,x_1,\ldots,x_k,x_k)$ and
$J=\diag(-J_1,+J_1,\ldots,-J_k,+J_k)$. In the sequel, we use the short
hand notations $x_p^\pm = x_p - iL_p\varepsilon$ and
$x^\pm=\diag(x_1^\pm,x_1^\pm,\ldots,x_k^\pm,x_k^\pm)$.  The product of
the differentials of all independent matrix elements is the volume
element $d[H]$. We use the notation and the conventions of
Refs.~\cite{TG,Guh96,GUH4}. The normalization $Z_k(x)=1$ at $J=0$
follows immediately from the definition~(\ref{gen}).

We wish to study whether the generating function can be represented as
an integral of the form
\begin{equation}
Z_k(x+J) = \int d[\sigma]  \, Q(\sigma) \detg^{-N}\left(\sigma-x^\pm-J\right) \ ,
\label{gents}
\end{equation}
where $\sigma$ is a $2k\times 2k$ supermatrix with Hermitean or
related symmetries, and where $\detg$ denotes the superdeterminant. If
such a representation can be shown to exist, the question arises
whether the probability density $Q(\sigma)$ in superspace can be
obtained in a unique way from the probability density $P(H)$ in
ordinary space.

\subsection{Rotation Invariant Probability Densities}
\label{sec22}

For the important class of norm--dependent ensembles, i.e.~ensembles
defined by a probability density depending exclusively on $\tr H^2$,
such a unique construction is indeed possible and was performed in
Ref.~\cite{TGnorm}. Here, we tackle the problem of arbitrary rotation
invariant probability densities $P(H)$. We recall that a probability
density must be normalizable and positively semi--definite. The term
``arbitrary'' has to be understood as excluding those functions $P(H)$
which would lead to a divergent integral~(\ref{gen}).  By ``rotation
invariant'' we mean that the probability density has the property
\begin{equation}
P(H) = P(U_0HU_0^\dagger) = P(E) \ , 
\label{inv}
\end{equation}
where $U_0$ is any fixed matrix in ${\rm SU}(N)$ and where
$E=\diag(E_1,\ldots,E_N)$ is the diagonal matrix of the eigenvalues of
$H$. Although it is obvious, we underline that this includes
invariance under permutations of the vectors defining the basis in
which $H$ is written down and also invariance under permutations of
the eigenvalues. Hence, the probability density $P(H)$ should depend
only on matrix invariants, such as $\tr H^m$ where $m$ is real and
positive.  Anticipating the later discussion, we already now mention
that this requirement is a most natural one in view of the general
character of the supersymmetry method. The strength of this method is
rooted in the drastic reduction of degrees of freedom, i.e.~of the
number of integration variables, when an integral over the $N\times N$
matrix $H$ is identically rewritten as an integral over the $2k \times
2k$ matrix $\sigma$. Thus, supersymmetry removes a certain redundancy.
The rotation invariance requirement implies precisely this redundancy
which the supersymmetry method needs. We will show that this holds for
arbitrary rotation invariant probability densities $P(H)$.

\subsection{Mutual Relations between the Different Correlation Functions}
\label{sec23}

We wish to address the correlation functions~(\ref{corrdefi})
and~(\ref{corrdef}) for finite level number $N$, we are not aiming at
an asymptotic discussion. If a saddlepoint approximation leading to a
non--linear $\sigma$ model as in Refs.~\cite{EFE83,VWZ,SW80} is the
method of choice to study a certain physics problem, one performs
precisely such an asymptotic expansion in $1/N$. This is not what we
are going to do in the present contribution.  Admittedly, our goal to
address the problem exactly for finite $N$ renders our task
mathematically demanding, because we have to solve certain group
integrals. One the other hand, luckily and at first sight
paradoxically, this goal allows us to circumvent the introduction of
hyperbolic symmetry, which is a deeply rooted, non--trivial feature of
the non--linear $\sigma$ model~\cite{EFE83,VWZ,ZIR04,SW80}. Hyperbolic
symmetry means that the ensuing supersymmetric representation of the
random matrix model must involve non--compact groups to make the
integrals convergent. This is inevitable if the imaginary increments
of the energies lie on different sides of the real axis.  However, if
they lie on the same side, no hyperbolic symmetry occurs and all
groups are compact.  This facilitates the supersymmetric treatment
tremendously.

We now argue that the correlation functions~(\ref{corrdefi}) of the
imaginary parts can be recovered from the more general correlation
functions~(\ref{corrdef}) that are suited for the supersymmetric
treatment, even if all imaginary increments lie on the same side of
the real axis. We choose $L_p=+1$ for all $p=1,\ldots,k$.  Upon
Fourier transforming the correlation functions~(\ref{corrdef}), we
obtain the $k$--point correlations in the domain of the times $t_p , \
p=1,\ldots,k$,
\begin{eqnarray} 
\widehat{r}_k(t_1,\ldots,t_k) &=& \frac{1}{\sqrt{2\pi}^k}
           \int\limits_{-\infty}^{+\infty} dx_1 \exp\left(it_1x_1\right) \cdots
           \int\limits_{-\infty}^{+\infty} dx_k \exp\left(it_kx_k\right)                      
                                                          \nonumber\\ 
 &  &   \qquad\qquad\qquad\qquad\qquad
           \widehat{R}_k(x_1,\ldots,x_k)
                                                          \nonumber\\ 
 &=&   \left(i2\right)^k
                  \prod_{p=1}^k \Theta(t_p)\exp\left(-\varepsilon t_p\right)  \, r_k(t_1,\ldots,t_k)
\label{tcorr}
\end{eqnarray}
with 
\begin{equation}
r_k(t_1,\ldots,t_k) = \frac{1}{\sqrt{2\pi}^k}
                            \int d[H] \, P(H) \prod_{p=1}^k \tr \exp\left(iHt_p\right) \ .
\label{tcorri}
\end{equation}
Importantly, this latter $k$--point correlation function
$r_k(t_1,\ldots,t_k)$ in time domain is precisely the Fourier
transform of the correlation function~(\ref{corrdefi}).  It is
well--defined on the entire real axes of all its arguments $t_p$. The
inverse transform yields
\begin{eqnarray}
R_k(x_1,\ldots,x_k) &=& \frac{1}{\sqrt{2\pi}^k}
           \int\limits_{-\infty}^{+\infty} dt_1 \exp\left(-ix_1t_1\right) \cdots
           \int\limits_{-\infty}^{+\infty} dt_k \exp\left(-ix_kt_k\right)
                              \nonumber\\ 
                                 &  &  \qquad\qquad\qquad\qquad\qquad r_k(t_1,\ldots,t_k) \ .
\label{tcorrifou}
\end{eqnarray}
Based on this observation, we will pursue the following strategy in
later Sections of this contribution: We perform exact manipulations of
the correlations $\widehat{R}_k(x_1,\ldots,x_k)$ with $L_p=+1 , \
p=1,\ldots,k$, or of their generating functions, respectively. Having
obtained the appropriate supersymmetric representation, we Fourier
transform it into the time domain and find
$\widehat{r}_k(t_1,\ldots,t_k)$. In this expression, we then identify
the supersymmetric representation of the correlation functions
$r_k(t_1,\ldots,t_k)$. Upon backtransforming we arrive at the desired
supersymmetric representation for the correlation functions
$R_k(x_1,\ldots,x_k)$.

We can even extend the line of arguing. Once we have found the
supersymmetric representation of $r_k(t_1,\ldots,t_k)$, we can
construct the one of $\widehat{R}_k(x_1,\ldots,x_k)$ for any arbitrary
choice of the quantities $L_p=\pm 1$ by calculating
\begin{eqnarray}
\widehat{R}_k(x_1,\ldots,x_k) &=& \frac{1}{\sqrt{2\pi}^k}
           \int\limits_{-\infty}^{+\infty} dt_1 \exp\left(-ix_1t_1\right) \cdots
           \int\limits_{-\infty}^{+\infty} dt_k \exp\left(-ix_kt_k\right)
                              \nonumber\\ 
                                 &  &  \quad
                 \prod_{p=1}^k \left(i^{L_p}2\right)\Theta(L_pt_p) 
                             \exp\left(-\varepsilon L_pt_p\right) \, r_k(t_1,\ldots,t_k) \ .
\label{tcorrafou}
\end{eqnarray}
In this manner, we will obtain supersymmetric integral representations
for all correlation functions~(\ref{corrdef}) where the imaginary
increments lie independently of each other on either side of the real
axes, without introducing hyperbolic symmetry.

\section{Generalized Hubbard--Stratonovich Transformation}
\label{sec3} 

To carry out the program outlined in Section~\ref{sec2}, we have to
generalize the procedure referred to as Hubbard--Stratonovich
transformation accordingly.  In Section~\ref{sec31}, we Fourier
transform the probability density. An algebraic duality between matrix
structures is uncovered in Section~\ref{sec32}, and explored further
in Section~\ref{sec33}, where spectral decompositions of the matrices
involved are performed. Although our main interest are the correlation
functions where all $L_p$ are equal, we make these latter steps for an
arbitrary metric $L$. We do so, because we find it worthwhile to
document how natural the duality is even for a general
metric. Moreover, it allows us to clearly identify the point where a
general metric would require a much involved discussion of hyperbolic
symmetry --- which we then avoid by setting $L_p=+1$ for all
$p=1,\ldots,k$. In Section~\ref{sec34}, we construct the probability
density in superspace. We derive a generalized transformation formula
and the corresponding generating function in Sections~\ref{sec35}
and~\ref{sec36}, respectively. In Section~\ref{sec37}, the
norm--dependent ensembles are discussed as a simple example.

\subsection{Fourier Transform of the Probability Density}
\label{sec31}

The determinants in the generating function~(\ref{gen}) are written as
Gaussian integrals, those in the denominator as integrals over $k$
vectors $z_p , \ p=1,\ldots,k$ with $N$ complex commuting elements
each, and those in the numerator over $k$ vectors $\zeta_p , \
p=1,\ldots,k$ with $N$ complex anticommuting elements each. Again
omitting irrelevant phase factors, we have
\begin{eqnarray} 
Z_k(x+J) &=& \int d[H] \, P(H) \nonumber\\ 
 & &   \prod_{p=1}^k \int d[z_p] \exp\left(iL_pz_p^\dagger(H - x_p + iL_p\varepsilon + J_p)z_p\right)
                                                          \nonumber\\ 
 & &   \qquad 
          \int d[\zeta_p] \exp\left(i\zeta_p^\dagger(H - x_p + iL_p\varepsilon - J_p)\zeta_p\right) \ ,
\label{gentf1}
\end{eqnarray}
where $d[z_p]$ and $d[\zeta_p]$ denote the products of the independent
differentials. To ensure convergence of the integrals over the
commuting variables, the quantities $L_p$ are inserted in front of the
bilinear forms in the exponent. This is not needed in the integrals
over the anticommuting variables because they are always convergent.
We order the quantities $L_p$ in the metric tensor
$L=\diag(L_1,1,\ldots,L_k,1)$.  Using the identities
\begin{equation} 
z_p^\dagger H z_p = \tr Hz_pz_p^\dagger \quad {\rm and} \quad
\zeta_p^\dagger H \zeta_p =  - \tr H\zeta_p\zeta_p^\dagger \ ,
\label{idtr}
\end{equation}
the average over $H$ in Eq.~(\ref{gentf1}) can be written as the Fourier transform
\begin{equation}
\Phi(K) = \int d[H] \, P(H) \exp\left(i \tr H K\right)
\label{foup}
\end{equation}
of the probability density. The Fourier variable is the matrix
\begin{equation}
K = \sum_{p=1}^k L_p z_pz_p^\dagger - \sum_{p=1}^k \zeta_p\zeta_p^\dagger \ .
\label{fouv}
\end{equation}
The function $\Phi(K)$ is referred to as characteristic function. The
definition~(\ref{foup}) of the Fourier transform is the one mostly
used in the statistics literature. It guarantees that $\Phi(0)=1$,
directly reflecting the normalization of $P(H)$. The definition of the
Fourier transform in Section~\ref{sec23} follows the ``symmetric
convention'' in which the same factor of $1/\sqrt{2\pi}$ appears in
the transform and in its inverse.

Up to now, all steps were exactly as in the case of a Gaussian
probability density $P(H)$.  In the Gaussian case, one can now do the
integral~(\ref{foup}) explicitly and one obtains a Gaussian form for
the characteristic function $\Phi(K)$. Here we consider a general
rotation invariant $P(H)$. Of course, we must assume that the Fourier
transform exists, i.e.~that $P(H)$ is absolutely integrable or, even
better, that it is a Schwartz function. Absolute integrability is
guaranteed by the fact that $P(H)$ is a probability density, implying
that it is positively semi--definite and normalized. However, we also
must assume that the integrals over the vectors $z_p$ converge after
doing the Fourier transform. The integrals over the vectors $\zeta_p$
can never cause convergence problems. In the Gaussian case, all those
convergence issues have been carefully discussed in Ref.~\cite{SW80},
a recent review is given in Ref.~\cite{ZIR04}. In the general rotation
invariant case, we have no other choice than to implicitly exclude
those probability density $P(H)$ which would cause convergence
problems, assuming that all integrals in the sequel converge.  We will
come back to this point later.

It is easy to see that the rotation invariance of $P(H)$ also implies
the rotation invariance of $\Phi(K)$.  The matrix $K$ is Hermitean,
$K^\dagger=K$. This is so for all choices $L_p=\pm 1$ of the metric
elements. As the entries of $K$ are commuting variables, we may
conclude that $K$ can be diagonalized,
\begin{equation}
K = \widetilde{V}Y\widetilde{V}^\dagger \ ,
\label{diagk}
\end{equation}
where $\widetilde{V}$ is in ${\rm SU}(N)$ and where
$Y=\diag(Y_1,\ldots,Y_N)$ is the diagonal matrix containing the
eigenvalues $Y_n, \ n=1,\ldots,N$ of $K$. The rotation invariance of
$P(H)$ and the invariance of the measure $d[H]$ allows one to absorb
$V$ such that the characteristic function $\Phi(K)$ depends only on
$Y$,
\begin{equation}
\Phi(K) = \int d[H] \, P(H) \exp\left(i \tr H Y\right) = \Phi(Y) \ .
\label{foupi}
\end{equation}
In other words, $\Phi(K)$ is a rotation invariant function, too.

\subsection{Underlying Algebraic Duality}
\label{sec32}

The merit of the Hubbard--Stratonovich transformation in the
supersymmetry method is the drastic reduction in the number of degrees
of freedom. This is rooted in a duality between matrices in ordinary
and superspace. We uncover this duality and cast it into a form which
allows a straightforward generalization of the previous discussion for
a Gaussian probability density.  We define the $N \times 2k$
rectangular matrix
\begin{equation}
A = \left[ z_1 \cdots z_k \ \zeta_1 \cdots \zeta_k \right] \ .
\label{adef}
\end{equation}
Although it contains commuting and anticommuting entries, $A$ is not a
supermatrix of the type commonly appearing in the framework of the
supersymmetry method~\cite{EFE83,EFE97}.  Nevertheless, this matrix
will play a crucial r\^ole in the following.  Its Hermitean
conjugate is the $2k \times N$ rectangular matrix
\begin{equation}
A^\dagger = \left[\begin{array}{c} z_1^\dagger \\
                                                        \vdots \\
                                                        z_k^\dagger \\ 
                                                        -\zeta_1^\dagger \\
                                                        \vdots \\
                                                        -\zeta_k^\dagger       
                                                          \end{array}\right] \ .
\label{adefhc}
\end{equation}
The inclusion of the minus signs is necessary to be consistent with the
conventions in Refs.~\cite{VWZ,TG,Guh96,GUH4}. It ensures that we have
$(A^\dagger)^\dagger=A$.

We notice that the boson--fermion block notation~\cite{VWZ} is used in
the definition~(\ref{adef}), which differs from the $pq$ block
notation~\cite{VWZ} employed when defining the supermatrices $L$, $x$
and $J$, as well as implicitly $\sigma$ in Section~\ref{sec2}. In the
boson--fermion block notation, first all commuting and then all
anticommuting variables (or vice versa) are collected in a
supervector.  Hence, the supermatrices which linearly transform those
vectors consist of rectangular (in the present case $k\times k$)
blocks of commuting or anticommuting variables. The $pq$ notation is
obtained by simply reordering the basis. One collects the commuting
and anticommuting variables corresponding to each energy index
$p=1,\ldots,k$, such that every supermatrix is written as a $k\times
k$ ordinary matrix with $2\times 2$ supermatrix elements assigned to
each index pair $(p,q)$. While the latter notation was handy when
introducing the generating function in Section~\ref{sec2}, it is more
convenient for the present discussion to use the boson--fermion block
notation.  In particular, the metric then reads
$L=\diag(L_1,\ldots,L_k,1,\ldots,1)$.

The Hermitean $N\times N$ matrix $K$ defined in Eq.~(\ref{fouv}) can
be written as the matrix product
\begin{equation}
K = A L A^\dagger = (AL^{1/2}) \, (L^{1/2}A^\dagger) \ .
\label{ksp}
\end{equation}
There exists a natural dual matrix to $K$, found by interchanging the
order of the matrices in Eq.~(\ref{ksp}). It is the $2k\times 2k$
matrix
\begin{equation}
B = (L^{1/2}A^\dagger) \, (AL^{1/2}) =  L^{1/2} A^\dagger A L^{1/2} \ ,
\label{bdp}
\end{equation}
where 
\begin{equation}
A^\dagger A = \left[\begin{array}{cccccc} 
                               z_1^\dagger z_1& \cdots & z_1^\dagger z_k & 
                               z_1^\dagger \zeta_1& \cdots & z_1^\dagger \zeta_k  \\
                               \vdots & & \vdots & \vdots & & \vdots \\
                               z_k^\dagger z_1& \cdots & z_k^\dagger z_k & 
                               z_k^\dagger \zeta_1& \cdots & z_k^\dagger \zeta_k  \\
                               -\zeta_1^\dagger z_1& \cdots & -\zeta_1^\dagger z_k & 
                               -\zeta_1^\dagger \zeta_1& \cdots & -\zeta_1^\dagger \zeta_k  \\
                               \vdots & & \vdots & \vdots & & \vdots \\
                               -\zeta_k^\dagger z_1& \cdots & -\zeta_k^\dagger z_k & 
                               -\zeta_k^\dagger \zeta_1& \cdots & -\zeta_k^\dagger \zeta_k  
                                \end{array}\right] \ .
\label{ada}
\end{equation}
While $K=A L A^\dagger$ is an ordinary matrix, $A^\dagger A$ and $B =
L^{1/2} A^\dagger A L^{1/2}$ are supermatrices. Moreover, $K$ is
Hermitean for all choices of the metric $L$, i.e.~for every
combination $L_p=\pm 1$, but $B$ is in general not Hermitean because
some entries of the metric are imaginary, $L_p^{1/2}=i$.  The
supermatrix $A^\dagger A$, however, is Hermitean. 

Interestingly, the duality between the matrices $K$ and $B$ also
implies the equality of invariants involving the traces according to
\begin{equation}
\tr K^m = \trg B^m \ ,
\label{tra}
\end{equation}
for every non--zero, positive integer $m$. This generalizes the case
of a Gaussian probability density where the need to discuss this
equality occurs only for $m=2$.  As the equality is not completely
trivial due to the presence of anticommuting variables,
Eq.~(\ref{tra}) is proven in~\ref{AppA}.

\subsection{Eigenvalues and Eigenvectors of the Dual Matrices}
\label{sec33}

Our way of formulating the algebraic duality is most helpful for the
spectral decomposition in ordinary and superspace. We write the
eigenvalue equation for the matrix $K$ as
\begin{equation}
K V_n = Y_n V_n \ ,
\label{kev}
\end{equation}
with $N$ eigenvectors $V_n , \ n=1,\ldots,N$.  We will now construct
them in such a way that they are not identical to those given as the
columns $\widetilde{V}_n$ of the unitary matrix $\widetilde{V}$
introduced in the diagonalization~(\ref{diagk}). For our construction,
we employ the $2k$ component supervectors
\begin{equation}
w_n = \left[\begin{array}{c} w_{n11}\\ 
                                              \vdots \\ 
                                              w_{nk1}\\ 
                                              w_{n12}\\
                                              \vdots \\ 
                                              w_{nk2} 
                   \end{array}\right] \ .
\label{wdef}
\end{equation}
There are two distinct representations of these supervectors. In the
first one, the elements $w_{npj}$ are commuting if $j=1$ and
anticommuting if $j=2$, in the second representation it is the other
way around.  We make the ansatz
\begin{eqnarray}
V_n = A L^{1/2} w_n
        = \sum_{p=1}^k z_p  L_p^{1/2}  w_{np1} + 
                                       \sum_{p=1}^k \zeta_p  w_{np2}  
\label{eva}
\end{eqnarray}
for the eigenvectors. It is convenient to multiply the coefficients
$w_{npj}$ from the right to avoid some cumbersome signs if the
$w_{np2}$ are anticommuting and appear together with the vectors
$\zeta_p$. We plug the ansatz~(\ref{eva}) into the eigenvalue
equation~(\ref{kev}) and find
\begin{eqnarray} 
K V_n &=& A L A^\dagger  A L^{1/2} w_n = A L^{1/2} \, B w_n \nonumber\\ 
Y_n V_n &=& Y_n A L^{1/2} w_n =  A L^{1/2} \, Y_n w_n \ ,
\label{anrein}
\end{eqnarray}
which yields $A L^{1/2}(Bw_n-Y_nw_n)=0$. Hence, we conclude that the
eigenvalue equation
\begin{equation}
B w_n = Y_n w_n 
\label{bev}
\end{equation}
holds if the eigenvalue equation~(\ref{kev}) is valid and if the
eigenvectors $V_n$ have the form~(\ref{eva}). There is a duality: the
eigenvalues $Y_n$ of $K$ to the eigenvectors $V_n$ in the
form~(\ref{eva}) are also eigenvalues of $B$ to the eigenvectors
$w_n$. 

The fact that the eigenvectors $w_n$ of the supermatrix $B$ belong to
one distinct representation as discussed below Eq.~(\ref{wdef})
implies that there are two types of eigenvalues corresponding to these
representations. We denote the $k$ eigenvalues associated with the
first representation by $y_{p1}=Y_p , \ p=1,\ldots,k$ and the $k$
eigenvalues associated with the second one by $y_{p2}=Y_{k+p} , \
p=1,\ldots,k$, respectively. Moreover, not all eigenvectors $V_n$ of
$K$ can have the form~(\ref{eva}) if the vector $w_n$ is required to
be eigenvector of $B$ at the same time. This is so, because $K$ and
$B$ have different dimensions $N\times N$ and $2k\times 2k$,
respectively. In all applications of RMT and supersymmetry, the level
number $N$ is large, such that we may safely assume $N>2k$. The matrix
$B$ has $2k$ eigenvalues.  Thus, the duality uncovered above only
makes a statement about $2k$ out of the $N$ eigenvalues of
$K$. Importantly, the remaining eigenvalues of $K$ are zero, because
$K$ is built upon $2k$ dyadic matrices. Hence, we have
\begin{equation}
Y_n = \left\{\begin{array}{ll} y_{p1} & \quad {\rm for} \quad n=p , \ p=1,\ldots,k\\
                                               y_{p2} & \quad {\rm for} \quad n=p+k , \ p=1,\ldots,k\\
                                               0         & \quad {\rm for} \quad n=2k+1,\ldots,N
               \end{array}\right. \ ,
\label{evch}
\end{equation}
if $N>2k$.  As $K$ is an ordinary Hermitean matrix, we know that the
eigenvectors $V_n$ to the zero eigenvalues can be chosen orthogonal
with each other and with those to the non--zero eigenvalues. We order
the non--zero eigenvalues in the $2k \times 2k$ diagonal supermatrix
\begin{equation}
y = \diag(y_{11}, \ldots, y_{k1},  iy_{k2}, \ldots,  iy_{k2})
\label{ydef}
\end{equation}
in boson--fermion block notation.  The definition includes an
imaginary unit $i$ coming with all eigenvalues $y_{p2}$. This is done
for convenience, the motivation will become clear later.

As the presence of the anticommuting variables requires some care, the
line of reasoning given above is supplemented with some details
in~\ref{AppB}, including the relation between the eigenvectors $V_n$
and $\widetilde{V}_n$.

\subsection{Probability Density in Superspace}
\label{sec34}

The characteristic function $\Phi(K)$ of the probability density $P(H)$
is according to Eq.~(\ref{foupi}) rotation invariant,
$\Phi(K)=\Phi(Y)$. Furthermore, by virtue of the previous discussion we
may view it as function of the eigenvalues of the supermatrix $B$,
such that we arrive at the chain of equalities
\begin{equation}
\Phi(K) = \Phi(Y) = \Phi(y) = \Phi(B) \ .
\label{chain}
\end{equation}
This crucial observation identifies $\Phi$ as an invariant function in
two different spaces, in ordinary space depending on the $N\times N$
matrix $K$ and in superspace, depending on the dual $2k\times 2k$
matrix $B$. It is interesting to notice that, if $\Phi(K)$ is a
function of all invariants $\tr K^m$ with $m=1,2,3,\ldots$, we
may conclude from the equality~(\ref{tra}) the identity
\begin{equation}
\Phi(\tr K,\tr K^2,\tr K^3,\ldots) = \Phi(\trg B,\trg B^2,\trg B^3,\ldots) \ , 
\label{pid}
\end{equation}
implying that the form of $\Phi(K)$ as function of those invariants
fully carries over to superspace. Although we have no reason to doubt
that this also holds in the presence of invariants $\tr K^m$ with
non--integer $m$, we have no proof, because we had to assume integer
$m$ when deriving the equality~(\ref{tra}). Luckily, this is not
important in the sequel. All what really matters is the general
insight expressed by the chain of equalities~(\ref{chain}). It
includes all invariants, except the determinant $\det K$ which is
trivially excluded, since we know from the previous
Section~\ref{sec33} that $K$ has zero eigenvalues if $N>2k$. Thus, we
do not employ Eq.~(\ref{pid}) in the sequel, although we will refer to
it at one point for illustrative purposes.

We restrict the further discussion to the case that all imaginary
increments of the energies lie on the same side of the real
axis. Hence we choose the metric
\begin{equation}
L = +1_{2k} \ .
\label{lchoice}
\end{equation}
This choice implies that the supermatrix $B$ becomes Hermitean,
$B^\dagger=B$, and the symmetry group is ${\rm U}(k/k)$, the unitary
supergroup in $k$ bosonic and $k$ fermionic dimensions. For a general
metric, the corresponding relation reads $B^\dagger=LBL$. The symmetry
group is pseudounitary, i.e.~the matrices of the defining
representation satisfy $w^\dagger Lw=L$. This hyperbolic symmetry
involves non--compact degrees of freedom. The situation was analyzed
in detail in Refs.~\cite{SW80} and~\cite{EFE83} for the non--linear
$\sigma$ model in ordinary and in superspace, respectively. The
proper, convergence ensuring integration manifolds of the
Hubbard--Stratonovich fields, corresponding to the matrices $\sigma$
in the present case, was constructed. It seems to us that the Gaussian
form of the probability densities $P(H)$ in these investigations was
somehow important for this construction. Here, however, we study
arbitrary rotation invariant probability densities $P(H)$. We did not
succeed in extending the line of reasoning in Refs.~\cite{EFE83,SW80}
to such general $P(H)$, even though we strongly believe that this
should be possible. Nevertheless, this does not cause a problem in
view of what we are aiming at, because we can proceed as outlined in
Section~\ref{sec23}. All issues related to convergence can be dealt
with much easier if the choice~(\ref{lchoice}) is made. That this
works fine in the case of a Gaussian $P(H)$ was already demonstrated
in Ref.~\cite{TG}.

As $\Phi(K)$ is the characteristic function of $P(H)$ in ordinary
space, the chain of equalities~(\ref{chain}) naturally suggests to
interpret $\Phi(B)$ as a characteristic function in superspace. To
this end, we introduce a probability density $Q(\sigma)$ depending on
a $2k\times 2k$ supermatrix $\sigma$ whose Fourier transform is
$\Phi(B)$. However, there is a subtle point to which we have to pay
attention. The symmetries of $B$ dictate to a large extent what the
symmetries of $\sigma$ have to be. As $B$ is a Hermitean supermatrix,
$\sigma$ ought to be a Hermitean supermatrix as well. Nevertheless, a
Wick--type--of rotation was applied in the case of Gaussian
probability densities which provides all elements in the
fermion--fermion block of $\sigma$ with an imaginary unit
$i$~\cite{EFE83,VWZ}. This modification is needed to solve a
convergence problem, too. It makes the expression $\trg\sigma^2$
positive semi--definite, and thus the integrals over the Gaussian
probability density $Q(\sigma) \sim \exp(-\trg\sigma^2)$
convergent. As we want to include the Gaussian as a special case in
our considerations, we also introduce this Wick--type--of rotation in
the $2k\times 2k$ supermatrix $\sigma$. The entries of the matrix $B$
can be modified accordingly by multiplying the vectors $\zeta_p$
containing the anticommuting variables with factors of $\sqrt{i}$, if
one wishes, but we do not do that here.  The diagonalization of the
matrix $\sigma$ can be written as
\begin{eqnarray}
\sigma = u s u^\dagger 
\quad {\rm with} \quad
s = \diag(s_{11}, \ldots, s_{k1}, is_{12}, \ldots,  is_{k2}) \ ,
\label{sdef}
\end{eqnarray}
where all eigenvalues $s_{pj}$ are real. The Wick--type--of rotation
multiplies the eigenvalues $s_{p2}$ with an imaginary unit.  Thus, $u$
is in the unitary supergroup ${\rm U}(k/k)$, without any modification
of its matrix elements. We also introduce a $2k\times 2k$ supermatrix
$\rho$ with the same symmetries as $\sigma$ and with the
diagonalization
\begin{eqnarray}
\rho = v r v^\dagger  
\quad {\rm with} \quad
r = \diag(r_{11}, \ldots, r_{k1},  ir_{12}, \ldots,  ir_{k2}) \ ,
\label{rdef}
\end{eqnarray}
where $v$ is in the unitary supergroup ${\rm U}(k/k)$ as well.
Anticipating the definitions of the eigenvalue matrices $s$ and $r$,
we introduced the eigenvalue matrix $y$ of $B$ in the
form~(\ref{ydef}).

We now define the probability density $Q(\sigma)$ in superspace
through the Fourier integral
\begin{equation}
\int d[\sigma] \, Q(\sigma) \exp\left(i\trg\sigma r\right)= \Phi(r) \ ,
\label{pdqa}
\end{equation}
or, as $\Phi(r)$ is invariant, we have equivalently
\begin{equation}
\int d[\sigma] \, Q(\sigma) \exp\left(i\trg\sigma\rho\right)= \Phi(\rho) \ ,
\label{pdqb}
\end{equation}
where $\Phi(\rho)$ is obtained from $\Phi(B)$ by formally replacing
$B$ with $\rho$.  We recall that the matrix elements of $B$ in the
fermion--fermion block are the scalar products
$\zeta_p^\dagger\zeta_q$ and thus nilpotent variables. This implies
that $\Phi(B)$ as a function of these variables is a finite power
series.  However, when replacing $B$ with $\rho$ we continue this
power series to an infinite one. It is important to realize that this
step is not problematic at all, because $\Phi(B)=\Phi(K)$ results from
the Fourier transform of $P(H)$. To illustrate the feasibility of this
continuation, we refer to the cases covered by Eq.~(\ref{pid}).

When writing out the expression $\trg\sigma\rho$, one sees that the
imaginary units due to the Wick--type--of rotation in the
fermion--fermion blocks nicely combine to $-1$ such that the whole
expression $\trg\sigma\rho$ is real. This is of course necessary to
make the Fourier transform well--defined.  The inverse of the Fourier
transform~(\ref{pdqb}) reads
\begin{equation}
Q(\sigma) = 2^{2k(k-1)} \int d[\rho] \, \Phi(\rho) \exp\left(-i\trg\sigma\rho\right) \ .
\label{pdsi}
\end{equation}
We notice that the prefactor $2^{2k(k-1)}$ does not involve $\pi$,
because we have the same number of commuting and anticommuting
variables.  Due to the invariance of the measure $d[\rho]$, the
rotation invariance of the characteristic function $\Phi(\rho)$ gives
with Eq.~(\ref{pdsi}) directly the same property for the probability
density, such that
\begin{equation}
Q(\sigma) = Q(s) \ .
\label{qinv}
\end{equation}
The rotation invariance of $P(H)$ implies the corresponding feature
for $Q(\sigma)$.  

There is a good reason why we defined $Q(\sigma)$ as above.
Nevertheless, what we need now to carry through our construction, is
the integral representation
\begin{equation}
\int d[\sigma] \, Q(\sigma) \exp\left(i\trg\sigma B\right)= \Phi(B) 
\label{pds}
\end{equation}
of the characteristic function considered as a function of $B$.  At
first sight, there is a problem, because $\trg\sigma B$ is not real
anymore. As the imaginary unit is present in the fermion--fermion
block of $\sigma$, but absent in that of $B$, the Fourier
integral~(\ref{pds}) seems ill--defined. However, as argued above, one
can also Wick--rotate the relevant elements of $B$.  Even if one
chooses not to do that, everything is under control, because the
matrix elements $\zeta_p^\dagger\zeta_q$ of $B$ in the
fermion--fermion block are nilpotent. The corresponding expressions in
$\exp\left(i\trg\sigma B\right)$ consist of a finite number of terms,
and no convergence problem for the $\sigma$ integration can occur.

\subsection{Generalized Transformation Formula}
\label{sec35}

After these preparations, we are in the position to derive a
transformation formula which expresses the probability density in
superspace as an integral over the probability density in ordinary
space. Using the result~(\ref{evch}), we have
\begin{eqnarray}
\tr H Y &=& \sum_{n=1}^N H_{nn}Y_n
                               \nonumber\\ 
            &=& \sum_{p=1}^k H_{pp}y_{p1} 
                           - \sum_{p=1}^k \left(iH_{(k+p)(k+p)}\right)\left(iy_{p2}\right) 
               = \trg h y 
\label{trhy}
\end{eqnarray}
with
\begin{equation}
h = \diag(H_{11}, \ldots, H_{kk},  iH_{(k+1)(k+1)}, \ldots, iH_{(2k)(2k)}) \ .
\label{hdef}
\end{equation}
Hence, Eq.~(\ref{foupi}) yields
\begin{equation}
\Phi(B) = \Phi(K) = \int d[H] \, P(H) \exp\left(i \trg h y\right) \ .
\label{foupihy}
\end{equation}
A proper definition of the diagonal matrix $h$ made it possible to
employ the supertrace in Eqs.~(\ref{trhy}) and~(\ref{foupihy}).  We
plug the characteristic function into the inverse Fourier
transform~(\ref{pdsi}) and find
\begin{equation}
Q(\sigma) = 2^{2k(k-1)} \int d[\rho] \exp\left(-i\trg\sigma\rho\right) \int d[H] \, 
                                        P(H) \exp\left(i \trg h r\right) \ ,
\label{trafo}
\end{equation}
where we use that $y$ and $r$ have precisely the same form.  Assuming
that the order of integrations may be interchanged, we arrive at the
generalized transformation formula
\begin{equation}
Q(\sigma) = \int d[H] \, P(H) \chi(\sigma,h) \ .
\label{trafochi}
\end{equation}
The function
\begin{equation}
\chi(\sigma,h) = 2^{2k(k-1)} \int d[\rho] \exp\left(i\trg(hr - \sigma\rho)\right) \ .
\label{chidef}
\end{equation}
is a projector which is related to, but different from a $\delta$
function. It might look surprising that the integrand contains the
full matrix $\rho$ as well as its eigenvalue matrix $r$, but recalling
the derivation, this is rather natural. The term $\exp(i\trg hr)$
stems from the Fourier transform of the probability density $P(H)$ in
ordinary space. Although it is conveniently written in a
supersymmetric notation, it is exclusively rooted in ordinary
space. Thus, anticommuting variables may only implicitly be present,
which makes it plausible that $r$ appears, but not the full $\rho$.

The projector satisfies the important normalization property
\begin{equation}
\int d[\sigma] \, \chi(\sigma,h) = \int d[\rho] \, \delta^{(4k^2)}(\rho) \exp\left(i\trg h r\right) 
                                                  = 1 \ ,
\label{chinorm}
\end{equation}
where $\delta^{(4k^2)}(\rho)$ is the product of the $\delta$ functions
of all $4k^2$ independent matrix elements in the supermatrix $\rho$.
This then gives directly the normalization
\begin{equation}
\int d[\sigma] \, Q(\sigma) = \int d[H] \, P(H) \int d[\sigma] \, \chi(\sigma,h) 
                                           = \int d[H] \, P(H) = 1
\label{qnorm}
\end{equation}
of the probability density in superspace. As one should expect, the
normalization of $P(H)$ yields the normalization of $Q(\sigma)$. One
is tempted to conclude that this feature wraps up the whole
convergence discussion if the choice~(\ref{lchoice}) has been
made. Such an interpretation is corroborated by the character of the
projector $\chi(\sigma,h)$ which is truly convergence friendly under
the integral. Nevertheless, this thinking comes to terms when
considering the complexity of all intermediate steps. Unfortunately,
it prevents us at present from providing the impression stated above
with more mathematical substance for a general $P(H)$. One possible
problem is related to the Wick--type--of rotation. All invariants $\tr
H^{2m}=\tr E^{2m}$ are positive semi--definite for all integer
$m$. This is clearly not so for the corresponding invariants
$\trg\sigma^{2m}=\trg s^{2m}$, where we have positive
semi--definiteness only for odd integers $m$.  This does not
inevitably lead to difficulties, because a term $\exp(-\tr H^{2m})$ in
$P(H)$ is not necessarily mapped onto its analog
$\exp(-\trg\sigma^{2m})$ in $Q(\sigma)$, but it illustrates at which
points problems could arise. Nevertheless, anticipating the discussion
to follow in Sections~\ref{sec4} and~\ref{sec5}, we mention already
now that the whole problem can be considered exclusively in Fourier
superspace such that only the convergence properties of the
characteristic function matter.

\subsection{Generating Function}
\label{sec36}

Having obtained the probability density $Q(\sigma)$, we use
Eqs.~(\ref{chain}) and~(\ref{pds}) in formula~(\ref{gentf1}). The
remaining steps to be done are then exactly as in Ref.~\cite{TG}, and
we arrive at the result
\begin{equation}
Z_k(x+J) = \int d[\sigma]  \, Q(\sigma) \detg^{-N}\left(\sigma-x^--J\right) \ ,
\label{gentsf}
\end{equation}
where
\begin{equation}
Q(\sigma) = \int d[H] \, P(H) \chi(\sigma,h) 
\label{qtrafo}
\end{equation}
is the probability density in superspace.

\subsection{Norm--dependent Ensembles Revisited}
\label{sec37}

The transformation formula~(\ref{qtrafo}) generalizes a transformation
formula which we obtained for norm--dependent random matrix
ensembles~\cite{TGnorm}. We revisit this case to acquire some
experience with the generalized transformation formula. The
probability density $P(H)$ of a norm--dependent ensemble depends on
$H$ only via $\tr H^2$. In Ref.~\cite{MK05}, the class of these
ensembles was constructed by averaging Gaussian probability densities
over the variance $t$,
\begin{equation}
P(H) = \int\limits_0^\infty f(t) \, \frac{1}{2^{N/2}(\pi t)^{N^2/2}}
                               \exp\left(-\frac{1}{2t}\tr H^2\right) \, dt \ .
\label{spread}
\end{equation}
where the choice of the spread function $f(t)$ determines the
ensemble. With the transformation formula~(\ref{qtrafo}), we find
\begin{eqnarray}
Q(\sigma) &=& 2^{2k(k-1)} \int\limits_0^\infty dt \, f(t) 
                     \, \int d[\rho] \, \exp\left(-i\trg\sigma\rho\right)
                                        \nonumber\\
      & & \qquad\qquad \int d[H] \, \frac{1}{2^{N/2}(\pi t)^{N^2/2}}
                        \exp\left(-\frac{1}{2t}\tr H^2\right) \, \exp\left(i\trg hr\right)
                                        \nonumber\\
      &=& 2^{2k(k-1)} \int\limits_0^\infty dt \, f(t) 
                     \, \int d[\rho] \, \exp\left(-i\trg\sigma\rho\right) 
                                         \nonumber\\
      & & \qquad\qquad  \int d[h] \, \frac{1}{(2\pi t)^{k/2}}
                         \exp\left(-\frac{1}{2t}\trg h^2\right) \, \exp\left(i\trg hr\right)
                                        \nonumber\\
      &=& 2^{k(k-1)} \int\limits_0^\infty dt \, f(t) 
                     \, \int d[\rho] \, \exp\left(-i\trg\sigma\rho\right) \, 
                                 2^{k(k-1)} \exp\left(-\frac{t}{2}\trg r^2\right) 
                                        \nonumber\\
      &=& \int\limits_0^\infty dt \, f(t) \, 2^{k(k-1)} \exp\left(-\frac{1}{2t}\trg\sigma^2\right) \,
\label{qnd}
\end{eqnarray}
which is indeed the correct result. We mention in passing that it
allows one to express the mapping of norm--dependent ensembles from
ordinary to superspace as one single integral in terms of the
probability density alone~\cite{TGnorm}, i.e.~without explicit
appearance of the spread function.

\section{Supersymmetric Formulation in Fourier Superspace}
\label{sec4}

Another supersymmetric formulation of the generating function will
prove most helpful for calculations of the correlation functions later
on. Also from a conceptual viewpoint, it has some rather appealing
features. In Section~\ref{sec41}, we construct the new formulation by
exploiting a convolution theorem, and in Section~\ref{sec42} we give a
direct derivation.

\subsection{Applying a Convolution Theorem}
\label{sec41}

According to Eq.~(\ref{gentsf}), $Z_k(x+J)$ is a convolution in
supermatrix space. For three $2k\times 2k$ Hermitean supermatrices
$\sigma$, $\rho$, $\tau$ and for two well--behaved functions
$g_1(\sigma)$, $g_2(\sigma)$ as well as their Fourier
transforms $G_1(\rho)$, $G_2(\rho)$, one easily derives the
convolution theorem
\begin{equation}
\int d[\sigma] \, g_1(\sigma) g_2(\tau-\sigma) = 
          2^{2k(k-1)} \int d[\rho] \, \exp\left(-i\trg\tau\rho\right) G_1(\rho) G_2(\rho) \ .
\label{conv}
\end{equation}
In the present case, we have $\tau=x+J$. We already know the Fourier
transform of $Q(\sigma)$, it is just the characteristic function
$\Phi(\rho)$.  The Fourier transform
\begin{equation}
I(\rho) = \int d[\sigma] \, \exp\left(i\trg\rho\sigma\right) \detg^{-N}\sigma^- \ .
\label{foudetg}
\end{equation}
of the superdeterminant is needed. It can be viewed as a
supersymmetric generalization of the Ingham--Siegel integral, whose
ordinary version has recently been used in the framework of
supersymmetric methods~\cite{YF}. Obviously, $I(\rho)$ only depends on
the eigenvalues $r$ of $\rho$. In~\ref{AppC}, we show that it is given
by
\begin{eqnarray}
I(\rho) &=& c_{Nk} 
                  \prod_{p=1}^k \Theta(r_{p1}) (ir_{p1})^N \exp\left(-\varepsilon r_{p1}\right)
                 \, \frac{\partial^{N-1}\delta(r_{p2})}{\partial r_{p2}^{N-1}} 
                                          \nonumber\\
c_{Nk} &=& \frac{1}{2^{k(k-1)}} \left(\frac{i2\pi (-1)^{N-1}}{(N-1)!}\right)^k \ .
\label{fdr}
\end{eqnarray}
We notice that $I(\rho)$ is almost equal to $\detg^{+N}\rho$, apart
from the restriction to negative eigenvalues $r_{p1}$ and the
occurrence of the functions $\delta(r_{p2})$ instead of $1/r_{p2}^\pm$.
Loosely speaking, the Fourier transform maps the superdeterminant
raised to the power $-N$ onto the superdeterminant raised to the power
$+N$. We find from Eqs.~(\ref{gentsf}) and~(\ref{conv})
\begin{equation}
Z_k(x+J) = 2^{2k(k-1)} \int d[\rho] \, \exp\left(-i\trg(x+J)\rho\right) \Phi(\rho) I(\rho) \ .
\label{gentconv}
\end{equation}
Thus we arrive at the remarkable insight that only the characteristic
function $\Phi(K)$ is needed in the generating function and, thus, for
the calculation of the correlation functions. It is of considerable
conceptual interest that the probability density in superspace
$Q(\sigma)$ follows in a unique way from the one in ordinary space
$P(H)$, but the use of $Q(\sigma)$ can be avoided if the Fourier
superspace representation is more convenient in a particular
application.

\subsection{Direct Derivation}
\label{sec42}

Since $Q(\sigma)$ does not appear anymore in the
expression~(\ref{gentconv}), the question arises if one can obtain
this result without going through the construction of the probability
density. This is indeed possible. We go back to Eq.~(\ref{gentf1}) and
do the average over the ensemble,
\begin{eqnarray} 
Z_k(x+J) &=& 
    \prod_{p=1}^k \int d[z_p] \exp\left(iL_pz_p^\dagger(iL_p\varepsilon -x_p + J_p)z_p\right)
                                                          \nonumber\\ 
 & &   \qquad 
          \int d[\zeta_p] \exp\left(i\zeta_p^\dagger(iL_p\varepsilon -x_p - J_p)\zeta_p\right) 
                                       \Phi(K) \ .
\label{ddz}
\end{eqnarray}
We now use the insights of Section~\ref{sec34} and insert an integral
over a $\delta$ function,
\begin{eqnarray}
\Phi(K) &=& \Phi(B) 
                                \nonumber\\ 
           &=& \int d[\rho] \, \Phi(\rho) \, \delta^{(4k^2)}(\rho-B)
                                                          \nonumber\\ 
           &=& 2^{2k(k-1)} \int d[\rho] \, \Phi(\rho) \, 
                                     \int d[\sigma] \exp\left(-i\trg\sigma(\rho-B)\right) \ ,
\label{pdb}
\end{eqnarray}
where $\rho$ and $\sigma$ are $2k \times 2k$ Hermitean supermatrices,
to which the Wick--type--of rotation has been applied in the
fermion--fermion blocks.  Again, one might argue that this makes the
expressions in Eq.~(\ref{pdb}) ill--defined, because these matrices
and the matrix $B$ are treated on equal footing, although no
Wick--type--of rotation has been applied to the latter. The same
reasoning as in Section~\ref{sec34} can be employed: Either one also
Wick--rotates $B$ or one argues that the integrals in Eq.~(\ref{pdb})
are well--defined because the elements of $B$ in the fermion--fermion
block are in any case nilpotent. We plug Eq.~(\ref{pdb}) into
Eq.~(\ref{ddz}). The integrals over the vectors $z_p$ and $\zeta_p$
can then be done in the usual way, and we have
\begin{eqnarray} 
Z_k(x+J) &=& 2^{2k(k-1)} \int d[\rho] \, \Phi(\rho) 
                                                         \nonumber\\ 
                & & \qquad\qquad \int d[\sigma] \, \exp\left(-i\trg\sigma\rho\right) \,
                                            \detg^{-N}\left(\sigma-x^--J\right)
                                                          \nonumber\\ 
                &=& 2^{2k(k-1)} \int d[\rho] \, \Phi(\rho) \, \exp\left(-i\trg(x+J)\rho\right)
                                                         \nonumber\\ 
     & & \qquad\qquad \int d[\sigma] \, \exp\left(-i\trg\sigma\rho\right)
                                                           \, \detg^{-N}\sigma^+ \ ,
\label{dds}
\end{eqnarray}
where we shifted $\sigma$ by $x+J$ in the last step. The remaining
$\sigma$ integral is, after changing variables from $\sigma$ to
$-\sigma$, precisely of the Ingham--Siegel type~(\ref{foudetg}) and we
obtain Eq.~(\ref{gentconv}).

Of course, the probability density $Q(\sigma)$ is somewhat hidden in
Eq.~(\ref{pdb}). However, to actually obtain it, one has to do the
$\rho$ integral, which would require an interchange with the $\sigma$
integration.  Avoiding the introduction of the probability density
$Q(\sigma)$ in the derivation sheds new light on the convergence
issues. If $P(H)$ is a Schwartz function, $\Phi(K)$ is a Schwartz
function as well and the convergence discussion can be exclusively
restricted to the Fourier superspace and to the properties of the
characteristic function when passing from ordinary space, i.e.~from
$\Phi(K)$, to superspace, i.e.~to $\Phi(B)$ and $\Phi(\rho)$.

\section{Correlation Functions in Terms of Eigenvalue Integrals}
\label{sec5} 

In Section~\ref{sec51}, we briefly review the integrals that we need
over the unitary group in ordinary and in superspace.  We derive a
first general result by identifying fundamental correlations in
Section~\ref{sec52}. In Section~\ref{sec53}, we carry out the
procedure outlined in Section~\ref{sec23} and obtain supersymmetric
integral representations of the correlation functions for arbitrary
positions of the imaginary increments. Another general result is given
in Section~\ref{sec54}, exclusively in terms of eigenvalue integrals.
In Section~\ref{sec55}, we discuss a probability density involving
higher order traces as an example.

\subsection{Eigenvalue--angle Coordinates and Group Integrals}
\label{sec51} 

The Hermitean random matrix is diagonalized according to
$H=UEU^\dagger$ with $E=\diag(E_1,\ldots,E_N)$ and with $U$ being in
${\rm SU}(N)$. The volume element in these coordinates reads
\begin{equation}
d[H] = \frac{\pi^{N(N-1)/2}}{N! \prod_{n=1}^{N-1} n!} \Delta_N^2(E) d[E] d\mu(U) \ ,
\label{volea}
\end{equation}
where we introduced the Vandermonde determinant
\begin{equation}
\Delta_N(E) = \det\left[E_n^{m-1}\right]_{n,m=1,\ldots,N} = \prod_{n<m} (E_n-E_m) \ .
\label{van}
\end{equation}
The invariant measure $d\mu(U)$ in Eq.~(\ref{volea}) is normalized to
unity. We will make use of the Harish-Chandra--Itzykson--Zuber
integral~\cite{HC,IZ}
\begin{equation}
\int d\mu(U) \exp\left(i\tr UEU^\dagger R\right) = 
                                        \frac{\prod_{n=1}^{N-1} n!}{i^{N(N-1)/2}}
                                        \frac{\det\left[\exp(iE_nR_m)\right]_{n,m=1,\ldots,N}}
                                               {\Delta_N(E)\Delta_N(R)} \ ,
\label{hciz}
\end{equation}
where $R=\diag(R_1,\ldots,R_N)$ is also a diagonal matrix. In
particular, we will employ this result for the case that $R_n=0$ for
$n>2k$. This can be obtained in various ways, as for example in
Ref.~\cite{PPW},
\begin{eqnarray}
& & \int d\mu(U) \exp\left(i\tr UEU^\dagger R\right) =  \prod_{n=N-2k+1}^{N-1}\frac{n!}{i^n}
                                 \nonumber\\
& & \quad \frac{\det\left[\exp(iE_nR_1) \  \cdots \ \exp(iE_nR_{2k}) \
                              1 \ E_n \ \cdots \ E_n^{N-2k-1} \right]_{n=1,\ldots,N}}
               {\Delta_N(E)\Delta_{2k}(\widetilde{R})\prod_{n=1}^{2k}R_{n}^{N-2k}} \ ,
\label{hcizlim}
\end{eqnarray}
where we write $\widetilde{R}=\diag(R_1,\ldots,R_{2k})$.

In superspace, the diagonalizations of the Hermitean supermatrices
$\sigma=usu^\dagger$ and $\rho=vrv^\dagger$ have already been
introduced in Eqs.~(\ref{sdef}) and~(\ref{rdef}). The volume element
$d[\rho]$ reads in eigenvalue--angle coordinates~\cite{TG}
\begin{equation}
d[\rho] = B_k^2(r) d[r] d\mu(v) \ ,
\label{voleas}
\end{equation}
where the function
\begin{equation}
B_k(r) = \frac{\Delta_k(r_1)\Delta_k(ir_2)}{\prod_{p<q}(r_{p1}-ir_{q2})}
           = \det\left[\frac{1}{r_{p1}-ir_{q2}}\right]_{p,q=1,\ldots,k} 
\label{vans}
\end{equation}
is the superspace equivalent of the Vandermonde determinant. The
supersymmetric analog~\cite{TG,GGT} of the
Harish-Chandra--Itzykson--Zuber integral is given by
\begin{eqnarray}
& & \int d\mu(v) \exp\left(i\trg vrv^\dagger s\right) =  \frac{i^k}{2^{k^2}\pi^k}
                                 \nonumber\\
& & \qquad\qquad \frac{\det\left[\exp(ir_{p1}s_{q1})\right]_{p,q=1,\ldots,k}
                                                \det\left[\exp(ir_{p2}s_{q2})\right]_{p,q=1,\ldots,k}}
                                               {B_k(r)B_k(s)} \ .
\label{hcizs}
\end{eqnarray}
As in Refs.~\cite{Guh96,GUH4}, the normalization of the invariant
measure $d\mu(v)$ is chosen such that formula~(\ref{hcizs}), when
applied to a shifted Gaussian distribution, yields the proper $\delta$
function in the curved space of the eigenvalues for vanishing variance.

\subsection{General Result as an Average over the Fundamental Correlations}
\label{sec52} 

The supergroup integral~(\ref{hcizs}) can now directly be applied to
the Fourier superspace formulation~(\ref{gentconv}), because both of
the functions $\Phi(\rho)$ and $I(\rho)$ depend only on the
eigenvalues $r$. This is the merit compared to the original superspace
formulation~(\ref{gentsf}), to which the result~(\ref{hcizs}) cannot
be applied in general. In the case of a Gaussian probability density,
a shift of the integration matrix $\sigma$ by $x+J$ gives a form
suited for the application of the supergroup
integral~\cite{TG,Guh96,GUH4}. In the general case, however,
Eq.~(\ref{gentconv}) is much more convenient. We find
\begin{eqnarray}
& & Z_k(x+J) = 1 \, + \,  \frac{2^{k(k-1)}}{B_k(x+J)} \left(\frac{i}{2\pi}\right)^k 
                                      \nonumber\\
& & \qquad\qquad \int d[r] \, B_k(r) \exp\left(-i\trg(x+J)r\right) \Phi(r) I(r) \ .
\label{gres}
\end{eqnarray}
Two remarks are in order. The first term, i.e.~unity, stems from a
certain boundary contribution which only appears in superspace.  In
physics, it is often referred to as Efetov--Wegner--Parisi--Sourlas
term~\cite{EFE83,PaSou,Weg,ConGro}, while it goes by the name
Rothstein contribution~\cite{ROT} in mathematics. In the present case,
it yields the normalization $Z_k(x)=1$ of the generating function,
because one easily sees that $1/B_k(x+J)$ vanishes at $J=0$. Formally,
the boundary contribution is obtained by putting $\rho=0$ in the
integral~(\ref{gentconv}), by using $\Phi(0)=1$ and
$I(0)=1/2^{k(k-1)}$ according to Eq.~(\ref{fdr}) and to~\ref{AppC},
and by finally dividing the result with the factor $2^{k(k-1)}$ which
is due to our definition of the volume element $d[\rho]$. There are
various methods to explicitly justify this procedure in the case
$k=1$. In Ref.~\cite{Guh93}, for example, it is directly constructed
from Rothstein's theorem. However, there is a problem, because none of
those explicit methods could be extended so far to our
eigenvalue--angle coordinates for $k>1$. We can thus not exclude that
further boundary contributions exist. Nevertheless, as to be discussed
below, we are confident that they are not important for our purposes.

The second remark concerns the determinants in the
formula~(\ref{hcizs}) which are not present in Eq.~(\ref{gres}). As
the functions $\Phi(r)$ and $I(r)$ are invariant under permutations of
the variables $r_{p1}$ as well as of the variables $r_{p2}$, it
suffices to keep only one term of each determinant, because all others
yield the same under the integral. 

We can now proceed in different ways. Here, we begin with inserting
the characteristic function in the form
\begin{equation}
\Phi(r) = \int d[H] \, P(H) \exp\left(i\trg hr\right) 
\label{chfr}
\end{equation}
as given in Section~\ref{sec35}. Upon interchanging the $r$ and the
$H$ integral we find the expression
\begin{equation}
Z_k(x+J) = 1 \,  +  \,  \frac{(-\pi)^k}{B_k(x+J)}  \int d[H] \, P(H) 
                                        \widehat{R}_k^{\rm (fund)}(x+J-h) \ ,
\label{gresh}
\end{equation}
where we introduced the fundamental correlation function
\begin{equation}
\widehat{R}_k^{\rm (fund)}(s) = 
                         2^{k(k-1)} \int d[r] \, B_k(r) \exp\left(-i\trg s r\right) I(r)
\label{corrfund}
\end{equation}
as a new object. In Eq.~(\ref{gresh}), we have to set $s=x+J-h$. We
refer to the correlation function~(\ref{corrfund}) as fundamental, for
it gives all structural information about the correlations before
averaging over the probability density $P(H)$. The fundamental
correlation function is the Fourier transform of the function $I(r)$
in the curved eigenvalue space. It is closely related to the
backtransform of $I(\rho)$, i.e.~to the superdeterminant
$\detg^{-N}\sigma^-$, but it is not quite the same. We discuss that 
in~\ref{AppD}.

The result~(\ref{gresh}) is not a trivial reformulation of
Eq.~(\ref{gen}) defining the generating function. While it is obvious
from Eq.~(\ref{gen}) that only the $N$ eigenvalues of $H$ are relevant
for the ensemble average, Eq.~(\ref{gresh}) makes a different
statement, namely that only $2k$ diagonal elements of $H$ enter the
computation of the average. Using the determinant
structure~(\ref{vans}) of $B_k(r)$ and formula~(\ref{fdr}), we find
that the fundamental correlation function has the determinant
structure
\begin{equation}
\widehat{R}_k^{\rm (fund)}(s) = 
     \det\left[\widehat{C}^{\rm (fund)}(s_{p1},is_{q2})\right]_{p,q=1,\ldots,k} \ ,
\label{corrfundd}
\end{equation}
where the fundamental kernel is given by
\begin{eqnarray}
\widehat{C}^{\rm (fund)}(s_{p1},is_{q2}) &=& - \frac{(-1)^{N-1}}{\pi (N-1)!}
\int\limits_{-\infty}^{+\infty}\int\limits_{-\infty}^{+\infty} \frac{dr_1dr_2}{r_1-ir_2}
\exp\left(-i(r_1s_{p1}^++r_2s_{q2})\right) 
                          \nonumber\\
& & \qquad\qquad\qquad\qquad \Theta(r_1) (ir_1)^N 
                                      \frac{\partial^{N-1}\delta(r_2)}{\partial r_2^{N-1}} \ .
\label{kerfund}
\end{eqnarray}
We suppress the indices $p$ and $q$ in the integration variables $r_1$
and $r_2$.  It is shown in~\ref{AppD} that the fundamental kernel can
be written as
\begin{eqnarray}
\widehat{C}^{\rm (fund)}(s_{p1},is_{q2}) &=& 
- \frac{1}{\pi} \sum_{n=0}^{N-1} \frac{(is_{q2})^n}{n!} 
         \int\limits_0^\infty dr_1 (ir_1)^n \exp\left(-ir_1s_{p1}^-\right) 
                        \nonumber\\
&=& \frac{1}{\pi} \sum_{n=0}^{N-1} \frac{(is_{q2})^n}{(s_{p1}^-)^{n+1}} \ .
\label{kerfe}
\end{eqnarray}
As this is a finite geometric series, we may also write
\begin{equation}
\widehat{C}^{\rm (fund)}(s_{p1},is_{q2}) 
 = \frac{1}{\pi s_{p1}^-} \frac{1-(is_{q2}/s_{p1}^-)^N}{1-(is_{q2}/s_{p1}^-)} 
 = \frac{1}{\pi (s_{p1}^-)^N} \frac{(s_{p1}^-)^N-(is_{q2})^N}{s_{p1}^--is_{q2}} \ .
\label{kerfegs}
\end{equation}
The fact that the fundamental kernel has a representation as a finite
series and as a ratio of differences is reminiscent of and related to the
Christoffel--Darboux formula~\cite{Bate} in the theory of orthogonal
polynomials.

The correlation functions according to Eq.~(\ref{corr}) are then quickly obtained
using the steps of Ref.~\cite{TG}. We find
\begin{eqnarray}
\widehat{R}_k(x_1,\ldots,x_k) &=& \int d[H] \, P(H) \widehat{R}_k^{\rm (fund)}(x-h)
                        \nonumber\\
    &=& \int d[h] \, P^{\rm (red)}(h) \widehat{R}_k^{\rm (fund)}(x-h)
                        \nonumber\\
    &=& \int d[h] \, P^{\rm (red)}(h) 
                        \nonumber\\
    & & \
    \det\left[\widehat{C}^{\rm (fund)}(x_p-H_{pp},x_q-iH_{(k+q)(k+q)})\right]_{p,q=1,\ldots,k} \ .
\label{corrafc}
\end{eqnarray}
The correlation functions are convolutions of the the fundamental
correlations with the reduced probability density
\begin{equation}
P^{\rm (red)}(h) = \int d[H/h] \, P(H) 
\label{pred}
\end{equation}
found by integrating $P(H)$ over all variables except the $2k$
diagonal elements $h$ of $H$.  The result~(\ref{corrafc}) holds for
arbitrary rotation invariant Random Matrix Ensembles. We notice that
the reduced probability density is connected to the characteristic
function. One sees that either directly from Eq.~(\ref{chfr}) or by
performing the following steps,
\begin{eqnarray}
P^{\rm (red)}(h) &=& \int d[H^\prime] \, \delta(h^\prime-h) P(H^\prime) 
                        \nonumber\\
    &=& \frac{1}{(2\pi)^{2k}} \int d[r] \int d[H^\prime] \, 
                               \exp\left(i\trg (h^\prime-h)r\right) P(H^\prime) 
                        \nonumber\\
    &=& \frac{1}{(2\pi)^{2k}} \int d[r] \exp\left(-i\trg hr\right) \Phi(r) \ .
\label{redcha}
\end{eqnarray}
Hence, $P^{\rm (red)}(h)$ is the Fourier backtransform of the
characteristic function depending on the $2k$ coordinates $r$ ---
which are here viewed as describing a flat space --- onto a function
defined in the flat space with coordinates $h$.  This is very
different from Fourier transforms in curved space, when the
eigenvalues $r$ are interpreted as the radial coordinates of a
Hermitean supermatrix $\rho$.

It is somewhat surprising that the probability densities in ordinary
space $P(H)$ or $P^{\rm (red)}(h)$, respectively, suffice to write
down Eq.~(\ref{corrafc}). One might conclude that this obliterates the
above convergence discussion related to the functional forms of the
probability densities in ordinary and superspace and of the
characteristic function. Formula~(\ref{corrafc}) indeed gives reason
to be optimistic. However, we recall that the characteristic function
was used in the derivation, even though it does not appear any more
explicitly. Hence, we must require the existence of the characteristic
function and also that the $H$ and the $r$ integrations can be
interchanged when going from Eq.~(\ref{chfr}) to Eq.~(\ref{gresh}).

The inherent determinant structure~(\ref{corrfundd}) of the
fundamental correlations will be destroyed in general when averaging
over the Random Matrix Ensemble. It will be preserved if the reduced
probability density factorizes according to
\begin{equation}
P^{\rm (red)}(h) =  \prod_{p=1}^{2k} P^{\rm (diag)}(H_{pp}) \ .
\label{phfactor}
\end{equation}
However, this is not the only situation in which the determinant
structure survives. The Mehta--Mahoux theorem~\cite{Mehta} implies
that the correlation functions $R_k(x_1,\ldots,x_k)$ can be written
as determinants for all rotation invariant probability densities which
factorize in their eigenvalue dependence,
\begin{equation}
P(H) = P(E) = \prod_{n=1}^N P^{\rm (ev)}(E_n) \ .
\label{pefactor}
\end{equation}
One would not expect that the the factorizations~(\ref{phfactor})
or~(\ref{pefactor}) are completely independent, but we have not looked
into this further.  In the present context, it is more important that
the applicability of the Mehta--Mahoux theorem is limited to precisely
the case when the factorization~(\ref{pefactor}) holds. It is thus a
quite attractive feature of the result~(\ref{corrafc}) that it is
valid for all rotation invariant probability densities which have the
property $P(H) = P(E)$, but which do not need to have any
factorization property as in Eqs.~(\ref{phfactor})
or~(\ref{pefactor}). In this sense, formula~(\ref{corrafc}) is more
general than the Mehta--Mahoux theorem.  In Section~\ref{sec54} we
will give another result, also valid for all rotation invariant
probability densities. Since it is formulated in terms of integrals
over the eigenvalues, its structure is somewhat different from
formula~(\ref{corrafc}). 

As an easy check of our findings, we show in~\ref{AppE} that
Eq.~(\ref{corrafc}) yields immediately the GUE correlation
functions. This is important, because it strengthens our confidence
that we treated the Efetov--Wegner--Parisi--Sourlas
term~\cite{EFE83,PaSou,Weg,ConGro} consistently.

\subsection{Correlations Functions of the Imaginary Parts and 
for Arbitrary Positions of the Imaginary Increments}
\label{sec53} 

As discussed in Section~\ref{sec21}, the correlation functions
$R_k(x_1,\ldots,x_k)$ as defined in Eq.~(\ref{corrdefi}) are the main
object of our interest.  We now construct integral representations for
them and, in addition, also for all correlation functions
$\widehat{R}_k(x_1,\ldots,x_k)$ as defined in Eq.~(\ref{corrdef}) for
arbitrary positions of the imaginary increments.  To avoid
introduction of hyperbolic symmetry, we restricted ourselves from
Section~\ref{sec34} on to the case that all imaginary increments lie
on the same side of the real axis.  However, applying the strategy
outlined in Section~\ref{sec23}, we can recover every correlation
function that we want.  

It is convenient to use the general result~(\ref{corrafc}), allowing
us to conduct the construction by only looking at the fundamental
correlation function $\widehat{R}_k^{\rm (fund)}(x-h)$.  Due to its
determinant structure, it depends on one fixed energy $x_p$ either in
the form $\widehat{C}^{\rm (fund)}(x_p-H_{pp},x_p-iH_{(k+p)(k+p)})$ or
in the form $\widehat{C}^{\rm (fund)}(x_q-H_{qq},x_p-iH_{(k+p)(k+p)})
\widehat{C}^{\rm (fund)}(x_p-H_{pp},x_{q^\prime}-iH_{(k+q^\prime)(k+q^\prime)})$
where $q\neq p$ and $q^\prime\neq p$. From the first of the expressions~(\ref{kerfe})
we conclude that in both cases the dependence of $\widehat{R}_k^{\rm (fund)}(x-h)$
on the fixed energy $x_p$ is a finite sum of the terms
\begin{equation}
\widehat{\Lambda}_{nm}(x_p) = \left(iH_{(k+p)(k+p)}-x_p\right)^m
                \int\limits_0^\infty dr_1 (ir_1)^n \exp\left(-ir_1(x_p^--H_{pp})\right) \ , 
\label{omdef}
\end{equation}
where $n=m$ is possible. As the average over the ensemble is linear,
it suffices to investigate the functions $\widehat{\Lambda}_{nm}(x_p)$
in order to study the energy dependence of the correlation functions
$\widehat{R}_k(x_1,\ldots,x_k)$. According to Section~\ref{sec23}, we
study the Fourier transform
\begin{equation}
\widehat{\lambda}_{nm}(t_p) = \frac{1}{\sqrt{2\pi}}
                  \int\limits_{-\infty}^{+\infty} dx_p \exp\left(it_px_p\right) 
                                                  \widehat{\Lambda}_{nm}(x_p) \ .
\label{omfou}
\end{equation}
Shifting $x_p$ by $H_{pp}$, it can be cast into the form
\begin{eqnarray}
\widehat{\lambda}_{nm}(t_p) &=& \frac{1}{\sqrt{2\pi}} \exp\left(it_pH_{pp}\right) 
                  \int\limits_{-\infty}^{+\infty} dx_p \exp\left(it_px_p\right) 
                                       \nonumber\\
             & & \qquad
                      \left(iH_{(k+p)(k+p)}-H_{pp}-x_p\right)^m
                              \int\limits_0^\infty dr_1 (ir_1)^n \exp\left(-ir_1x_p^-\right)
                                       \nonumber\\
             &=& \frac{1}{\sqrt{2\pi}} \exp\left(it_pH_{pp}\right) 
                                     \left(iH_{(k+p)(k+p)}-H_{pp}+i\frac{\partial}{\partial t_p}\right)^m
                                       \nonumber\\
             & & \qquad\qquad
                     \int\limits_0^\infty dr_1 (ir_1)^n \exp\left(-\varepsilon r_1\right)
                     \int\limits_{-\infty}^{+\infty} dx_p \exp\left(ix_p(t_p-r_1)\right) 
                                       \nonumber\\
             &=& \sqrt{2\pi} \exp\left(it_pH_{pp}\right) 
                                     \left(iH_{(k+p)(k+p)}-H_{pp}+i\frac{\partial}{\partial t_p}\right)^m
                                       \nonumber\\
             & & \qquad\qquad\qquad 
                     \int\limits_0^\infty dr_1 (ir_1)^n \exp\left(-\varepsilon r_1\right)
                                                                                  \delta(t_p-r_1) \ .
\label{omfouf}
\end{eqnarray}
As the $r_1$ integration extends over the positive real axis only, the
integral is zero whenever $t_p<0$. All derivatives are zero as well in
this case, implying that the entire expression is proportional to
$\Theta(t_p)$ . For $t_p>0$, the integral yields $(it_p)^n
\exp(-\varepsilon t_p)$. All derivatives of the exponential function
give terms containing powers of $\varepsilon$ and thus vanish in the
limit $\varepsilon\to 0$. Here, we may assume that the $t_p$ integral
cannot yield bare singularities in $\varepsilon$. We can thus neglect
all these terms and write $\exp(-\varepsilon t_p)$ in front of the
entire expression. We find
\begin{equation}
\widehat{\lambda}_{nm}(t_p) = i2 \Theta(t_p) \exp\left(-\varepsilon t_p\right)
                                                                  \lambda_{nm}(t_p)
\label{omfouhs}
\end{equation}
where
\begin{equation}
\lambda_{nm}(t_p) =  \frac{\sqrt{2\pi}}{i2} \exp\left(it_pH_{pp}\right) 
                                     \left(iH_{(k+p)(k+p)}-H_{pp}+i\frac{\partial}{\partial t_p}\right)^m
                                                      (it_p)^n \ .
\label{omfouhss}
\end{equation}
Indeed, Eq.~(\ref{omfouhs}) directly implies expression~(\ref{tcorr})
and we can read off the desired integral representations. The function
$\lambda_{nm}(t_p)$ is recognized as Fourier transform of
\begin{eqnarray}
\Lambda_{nm}(x_p) &=& -i\left(iH_{(k+p)(k+p)}-x_p\right)^m
                \int\limits_{-\infty}^{+\infty} dr_1 (ir_1)^n \exp\left(-ir_1(x_p-H_{pp})\right) \ , 
                                                        \nonumber\\
                                  &=& -i(-1)^n 2\pi \left(iH_{(k+p)(k+p)}-x_p\right)^m
                                            \frac{\partial^n}{\partial x_p^n} \delta(x_p-H_{pp})
                                                        \nonumber\\
                                  &=&  -2n! \left(iH_{(k+p)(k+p)}-x_p\right)^m
                                               \im\frac{1}{(x_p^--H_{pp})^{n+1}} \ .
\label{omhdef}
\end{eqnarray}
Collecting everything, we arrive at
\begin{equation}
R_k(x_1,\ldots,x_k) = \int d[h] \, P^{\rm (red)}(h) R_k^{\rm (fund)}(x-h)
\label{corrafch}
\end{equation}
with the fundamental correlation function
\begin{equation}
R_k^{\rm (fund)}(s) =
    \det\left[C^{\rm (fund)}(s_{p1},is_{q2})\right]_{p,q=1,\ldots,k} 
\label{corrfundh}
\end{equation}
and the fundamental kernel
\begin{eqnarray}
C^{\rm (fund)}(s_{p1},is_{q2}) &=& 
- \frac{1}{2\pi} \sum_{n=0}^{N-1} \frac{(is_{q2})^n}{n!} 
         \int\limits_{-\infty}^{+\infty} dr_1 (ir_1)^n \exp\left(-ir_1s_{p1}\right) 
                        \nonumber\\
&=& \frac{1}{\pi} \sum_{n=0}^{N-1} (is_{q2})^n\im\frac{1}{(s_{p1}^-)^{n+1}} \ .
\label{kerfeh}
\end{eqnarray}
Hence one simply has to replace the singularities $1/(s_{p1}^-)^{n+1}$
everywhere with their imaginary parts. Tracing back these
considerations, we realize that all necessary modifications reside in
the $r_{p1}$ integrals and specifically in the function
$I(r)$. Replacing Eq.~(\ref{fdr}) with
\begin{eqnarray}
I(\rho) = \frac{1}{2^{k(k-1)}} \left(\frac{\pi (-1)^{N-1}}{(N-1)!}\right)^k
                  \prod_{p=1}^k (ir_{p1})^N 
                 \, \frac{\partial^{N-1}\delta(r_{p2})}{\partial r_{p2}^{N-1}} 
\label{fdrh}
\end{eqnarray}
is equivalent to the above discussed steps made to obtain
$R_k(x_1,\ldots,x_k)$.

With the help of formula~(\ref{tcorrafou}), it is now an easy
exercise to construct integral representations for the correlation
functions $\widehat{R}_k(x_1,\ldots,x_k)$ defined in
Eq.~(\ref{corrdef}) with arbitrary positions of the imaginary
increments. Formulae~(\ref{corrfundd}) and~(\ref{corrafc}) remain
valid if the fundamental kernel is replaced with
\begin{eqnarray}
\widehat{C}^{\rm (fund)}(s_{p1},is_{q2}) &=& 
\mp \frac{1}{\pi} \sum_{n=0}^{N-1} \frac{(is_{q2})^n}{n!} 
         \int\limits_0^\infty dr_1 (ir_1)^n \exp\left(\mp ir_1s_{p1}^\mp\right) 
                        \nonumber\\
&=& \frac{1}{\pi} \sum_{n=0}^{N-1} \frac{(is_{q2})^n}{(s_{p1}^\mp)^{n+1}} 
 = \frac{1}{\pi (s_{p1}^\mp)^N} \frac{(s_{p1}^\mp)^N-(is_{q2})^N}{s_{p1}^\mp-is_{q2}} \ ,
\label{kerfearb}
\end{eqnarray}
where the notation $s_{p1}^\mp$ indicates that the imaginary increment
is chosen according to $x_p^\pm = x_p - iL_p\varepsilon$.  In terms of
the function $I(r)$, this is equivalent to replacing Eq.~(\ref{fdr})
with
\begin{equation}
I(\rho) = c_{Nk} 
                  \prod_{p=1}^k \Theta(L_pr_{p1}) (ir_{p1})^N \exp\left(-L_p\varepsilon r_{p1}\right)
                 \, \frac{\partial^{N-1}\delta(r_{p2})}{\partial r_{p2}^{N-1}} \ .
\label{fdrarb}
\end{equation}
Thus, we obtain supersymmetric integral representations for all these
correlation functions without using hyperbolic symmetry. 

The insights just presented may be viewed as a more formal
justification of the procedure denoted by the operator symbol $\Im$ in
Refs.~\cite{TG,Guh96,GUH4}. We argued in these studies that the
generating functions satisfy a diffusion process. The diffusion
propagator contains no information about the positions of the
imaginary increments, this is exclusively contained in the initial
condition of the diffusion.  Moreover, the diffusion propagator is
nothing but the supersymmetric Harish-Chandra--Itzykson--Zuber
integral~(\ref{hcizs}) over the unitary supergroup, not involving any
non--compact degrees of freedom. This can be verified in an elementary
way by simply plugging it into the diffusion equation. Hence, one is
free to adjust the positions of the imaginary increments as needed,
which essentially defined the operator $\Im$.  We have now given
another justification. Nevertheless, it remains an interesting
mathematical question to also derive all that from group integrals
involving non--compact degrees of freedom.

\subsection{General Result in Terms of Eigenvalue Integrals}
\label{sec54} 

A further integral representation follows from Eq.~(\ref{gres}). We
take the derivatives with respect to the source variables as in
Ref.~\cite{TG} and in Section~\ref{sec52} and find
\begin{equation}
\widehat{R}_k(x_1,\ldots,x_k) = 2^{k(k-1)}
                                     \int d[r] \, B_k(r) \exp\left(-i\trg xr\right) \Phi(r) I(r) \ .
\label{rgres}
\end{equation}
The correlation functions $R_k(x_1,\ldots,x_k)$ as well as those for
arbitrary positions of the imaginary increments are obtained as in the
previous Section~\ref{sec53}, we simply have to replace $I(r)$
according to Eq.~(\ref{fdr}) by $I(r)$ according to Eqs.~(\ref{fdrh})
or~(\ref{fdrarb}), respectively. We expand the determinant $B_k(r)$ by
introducing the permutations $\omega$ of the indices $p=1,\ldots,k$
and write
\begin{eqnarray}
\widehat{R}_k(x_1,\ldots,x_k) &=& 2^{k(k-1)}c_{Nk} \sum_{\omega} (-1)^{j(\omega)}
                        \nonumber\\
& & \quad \int d[r] \, \Phi(r) \prod_{p=1}^k 
                    \frac{\exp\left(-ix_pr_{p1}-x_{\omega(p)}r_{\omega(p)2}\right)}
                           {r_{p1}-ir_{\omega(p)2}}
                        \nonumber\\
& & \quad \ \Theta(L_pr_{p1}) (ir_{p1})^N \exp\left(-L_p\varepsilon r_{p1}\right)
                 \, \frac{\partial^{N-1}\delta(r_{\omega(p)2})}{\partial r_{\omega(p)2}^{N-1}} \ ,
\label{rgresexp}
\end{eqnarray}
where $j(\omega)$ is the parity of the permutation $\omega$.  The
$\delta$ functions allow us to do the $k$ integrals over the variables
$r_{p2}$ immediately. We integrate by parts and use Leibnitz' rule
to work out the derivatives of products,
\begin{eqnarray}
& & \widehat{R}_k(x_1,\ldots,x_k) = (i2\pi)^k \sum_{\omega} (-1)^{j(\omega)}
                        \nonumber\\
& & \qquad \sum_{n_1=0}^{N-1} \frac{1}{n_1!}
                 \int\limits_{-\infty}^{+\infty} dr_{11}\, \Theta(L_1r_{11})
                 \exp\left(-ix_1r_{11}-L_1\varepsilon r_{11}\right) (-ir_{11})^{n_1} \cdots
                        \nonumber\\
& & \qquad\qquad \sum_{n_k=0}^{N-1} \frac{1}{n_k!}
                    \int\limits_{-\infty}^{+\infty} dr_{k1}\, \Theta(L_kr_{k1})
                    \exp\left(-ix_kr_{k1}-L_k\varepsilon r_{k1}\right) (-ir_{k1})^{n_k}
                        \nonumber\\
& & \qquad\qquad\qquad\qquad
        \frac{\partial^{\sum_{p=1}^kn_p}\exp\left(-\sum_{p=1}^k x_{\omega(p)}r_{p2}\right) \Phi(r)}
                           {\prod_{p=1}^k \partial r_{p2}^{n_p}} \Bigg|_{r_2=0} \ .
\label{rgrespi}
\end{eqnarray}
This result is valid for an arbitrary rotation invariant probability
density. The structure of this expression is quite different from the
one in Section~\ref{sec52}, where the correlation functions were found
to be a convolution of the reduced probability density with the
fundamental correlations.

It is instructive to see how the correlation functions can acquire a determinant
structure. An obvious feature leading to this would be a factorization
\begin{equation}
\Phi(r) = \prod_{p=1}^k \Phi^{\rm (ev)}(r_{p1})\Phi^{\rm (ev)}(r_{p2}) 
\label{charfac}
\end{equation}
of the characteristic function. We find immediately
\begin{equation}
\widehat{R}_k(x_1,\ldots,x_k) = 
     \det\left[\widehat{C}(x_p,x_q)\right]_{p,q=1,\ldots,k} 
\label{correvch}
\end{equation}
with the kernel
\begin{eqnarray}
\widehat{C}(x_p,x_q) &=&  \frac{i}{\pi} \sum_{n=0}^{N-1} \frac{1}{n!}
\frac{\partial^n \exp\left(-x_qr_2\right) \Phi^{\rm (ev)}(r_2)}
                                             {\partial r_2^n} \Bigg|_{r_2=0} 
                               \nonumber\\
      & & \ \int\limits_{-\infty}^{+\infty} dr_1\, \Theta(L_pr_1)
                 \exp\left(-ix_pr_1-L_p\varepsilon r_1\right) (-ir_1)^n \Phi^{\rm (ev)}(r_1) \ ,
\label{kerrevch}
\end{eqnarray}
where we suppress the indices $p$ and $q$ in the $r$ variables. We
notice that the GUE case is trivially recovered. We then have
$\Phi^{\rm (ev)}(r_{pj})=\exp(-r_{pj}^2/4)$ which combines in the
derivative expression with the exponential to the generating function
of the Hermite polynomials, and the integral yields the generalized
Hermite functions as given in~\ref{AppE}. It is conceivable that
mechanisms other than following from the factorization~(\ref{charfac})
can be identified that also lead to a determinant structure. However,
as the merit of Eq.~(\ref{rgrespi}) is its completely general
character and its independence of such factorizations and determinant
structures, we have not explored this issue further.

One can wonder whether it is helpful to integrate over the group ${\rm
SU}(N)$, i.e.~over the ordinary unitary matrix $U$ diagonalizing $H$,
before inserting the characteristic function $\Phi(r)$ in
formula~(\ref{rgrespi}). With the help of Eq.~(\ref{hcizlim}) we find
\begin{eqnarray}
\Phi(r) &=&  \frac{\pi^{N(N-1)/2}}{N! \prod_{n=1}^{N-1} n!} 
                       \int d[E] \Delta_N^2(E) P(E) 
                       \int d\mu(U) \exp\left(i\tr UEU^\dagger R\right)
                                 \nonumber\\
&=&  \frac{\pi^{N(N-1)/2}}{i^{(N-k)(2k-1)}
           N! \prod_{n=1}^{N-2k} n! \, \Delta_{2k}(r_1,r_2)\prod_{p=1}^{2k}(r_{p1}r_{q2})^{N-2k}}
                                 \nonumber\\
& & \int d[E] \Delta_N(E) P(E)
                                 \nonumber\\
& & \det\left[\exp(iE_nr_{11}) \  \cdots \ \exp(iE_nr_{k2}) \
                              1 \ E_n \ \cdots \ E_n^{N-2k-1} \right]_{n=1,\ldots,N} 
\label{chrff}
\end{eqnarray}
where we have to set $R_p = r_{p1} , \ R_{p+k} = r_{p2} , \
p=1,\ldots,k$. The eigenvalues $r_{p2}$ do not come with an imaginary
increment in the formula above. This is also important in
$\Delta_{2k}(r_1,r_2)$ which is the ordinary Vandermonde determinant
of the $2k$ variables $r_1$ and $r_2$.  As the whole integrand is
invariant under permutations of the eigenvalues $E_n$, we may replace
the determinant stemming from the group integration by the product of
its diagonal elements, because all other terms yield the same
result. However, from the resulting expression
\begin{eqnarray}
\Phi(r) &=&  \frac{\pi^{N(N-1)/2}}{i^{(N-k)(2k-1)}
           \prod_{n=1}^{N-2k} n! \, \Delta_{2k}(r_1,r_2)\prod_{p=1}^{2k}(r_{p1}r_{p2})^{N-2k}}
                                 \nonumber\\
& & \quad \int d[E] \Delta_N(E) P(E)
                                 \nonumber\\
& & \qquad \prod_{p=1}^k \exp\left(i(E_pr_{p1}+E_{k+p}r_{p2})\right)
                                               \prod_{n=2k+1}^{N} E_n^{n-2k-1}
\label{chrffd}
\end{eqnarray}
it is not immediately obvious anymore that its limit of vanishing
$r_{pj}$ remains finite, given by the normalization $\Phi(0)=1$.

We give the expressions~(\ref{chrff}) and~(\ref{chrffd}) mainly for
the sake of completeness, because they are not particularly useful in
their general form. Although the powers in the denominator are not
real singularities in Eqs.~(\ref{chrff}) and~(\ref{chrffd}), they
become truly singular, if one tries to exchange the order of
integrations and to do the $r$ integrations first.

\subsection{Ensembles Involving Higher Order Traces as an Example}
\label{sec55} 

As it might be helpful to illustrate our findings by an example, we
consider the probability density
\begin{equation}
P(H) = b_{M_1M_2} \left(\tr H^{M_1}\right)^{M_2} \exp\left(-\tr H^2\right) 
\label{ptram}
\end{equation}
for a fixed pair of integers $M_1,M_2=0,1,2,\ldots$. The constant
$b_{M_1M_2}$ ensures normalization. The Gaussian case is recovered for
$M_1=0$ or $M_2=0$. A few obviously meaningless cases have to be
excluded, such as the choice $M_1=M_2=1$, which makes the
normalization integral vanish. While the probability
density~(\ref{ptram}) is still in the norm--dependent class discussed
in Ref.~\cite{TGnorm} for $M_1=2$, it is not for other values of
$M_1$.  Importantly, the probability density~(\ref{ptram}) does not
factorize according to Eqs.~(\ref{phfactor}) or~(\ref{pefactor}). In
particular, this means that this Random Matrix Ensemble is not covered
by the Mehta--Mahoux theorem, although we do not exclude that is
possible with some efforts to extend the latter
properly. Formula~(\ref{corrafc}) provides a direct way to calculate
the correlation functions for such ensembles. However, as we aim at
addressing the conceptual issues in the present contribution, we
refrain from presenting the quite cumbersome expressions too
explicitly. We rather sketch the calculation briefly and infer what
kind of structure the correlations functions will acquire.  It is
obvious from definition~(\ref{pred}) that the reduced probability
density has the form
\begin{equation}
P^{\rm (red)}(h) =  \exp\left(-\trg h^2\right) \sum_{\{m\}} a_{\{m\}} S_{\{m\}}(h) \ ,
\label{predtram}
\end{equation}
where the $a_{\{m\}}$ are constants and where
\begin{equation}
S_{\{m\}}(h) = \sum_{\omega} \prod_{p=1}^{2k} H_{pp}^{m_{\omega(p)}}
\label{symf}
\end{equation}
are symmetric functions, i.e.~linear combinations of products
involving a set $\{m\}$ of integer exponents $m_p$, symmetrized by
summing over all permutations $\omega$ of the indices $p=1,\ldots,2k$.
The exponents $m_p$ are between zero and $M_1+M_2$ with the
restriction that their sum does not exceed $M_1+M_2$.  Thus, the
correlation functions are given by
\begin{equation}
\widehat{R}_k(x_1,\ldots,x_k) 
    =  \sum_{\{m\}} a_{\{m\}} \int d[h] \, \exp\left(-\tr h^2\right) S_{\{m\}}(h) 
                                                              \widehat{R}_k^{\rm (fund)}(x-h) \ .
\label{corrptram}
\end{equation}
Upon inserting Eq.~(\ref{symf}) and using the determinant structure of
the fundamental correlations, we obtain
\begin{equation}
\widehat{R}_k(x_1,\ldots,x_k) 
    =  \sum_{\{m\}} a_{\{m\}} \sum_{\omega}
 \det\left[\widehat{C}_{m_{\omega(p)}m_{\omega(k+q)}}(x_p,x_q)\right]_{p,q=1,\ldots,k} 
\label{corrptdet}
\end{equation}
where the kernel
\begin{equation}
\widehat{C}_{m_1m_2}(x_p,x_q) = 
 \frac{1}{\pi} \exp\left(-x_p^2\right) \sum_{n=0}^{N-1} \frac{1}{n!} 
         \widehat{\eta}_{nm_1}(x_p)\vartheta_{nm_2}(x_q) 
\label{kermo}
\end{equation}
has a structure formally similar to that of the GUE kernel. The
functions
\begin{eqnarray}
\widehat{\eta}_{nm_1}(x_p) &=& 
 \int\limits_{-\infty}^{+\infty} dH_{pp} \exp\left(-H_{pp}^2\right)H_{pp}^{m_1}
 \int\limits_0^\infty dr_1 (ir_1)^n \exp\left(\mp ir_1(x_p^\mp-H_{pp})\right) 
                                 \nonumber\\
\vartheta_{nm_2}(x_q) &=& 
 \int\limits_{-\infty}^{+\infty} dH_{(k+q)(k+q)}
                             \exp\left(-H_{(k+q)(k+q)}^2\right)H_{(k+q)(k+q)}^{m_2}
                                 \nonumber\\
& & \qquad\qquad\qquad\qquad\qquad
                             \left(x_q-iH_{(k+q)(k+q)}\right)^n
\label{etatheta}
\end{eqnarray}
can be written as finite weighted sums of the generalized Hermite
functions which are discussed in~\ref{AppE} and of the ordinary Hermite
polynomials, respectively. According to the result~(\ref{corrptdet}),
the correlation functions are linear combinations of determinants.

Alternatively, this calculation can be carried out using the results
of Section~\ref{sec54}.  It follows from the inverse of
formula~(\ref{redcha}) that the characteristic function has a form
very similar to the reduced probability density,
\begin{equation}
\Phi(r) =  \exp\left(-\frac{1}{4}\trg r^2\right) \sum_{\{m\}} \widetilde{a}_{\{m\}} S_{\{m\}}(h) 
\label{predchar}
\end{equation}
with new constants $\widetilde{a}_{\{m\}}$. With the help of
Eq.~(\ref{rgres}) or~(\ref{rgrespi}) this leads in a straightforward
manner to the above mentioned linear combinations of determinants.

\section{Summary and Conclusions}
\label{sec6}

We derived supersymmetric formulations for arbitrary rotation
invariant Random Matrix Ensembles. The construction is based on an
algebraic duality between ordinary and superspace which made it
possible to generalize the Hubbard--Stratonovich transformation. We
identified an integral transformation that involves a projector and
yields the probability density in superspace from the one in ordinary
space. However, we showed that despite the conceptual insights thereby
obtained, the theory can be formulated without using the probability
density in superspace. It turned out that it is possible and often
even better to work in Fourier space, because the characteristic
functions of the probability density have the same functional form in
ordinary and superspace.  At present, it appears to us that it is a
priori easier to analyze some convergence issues in Fourier
superspace, but to make more definite statements will require
additional work. It is not inconceivable, that manifestly invariant
theories can be constructed in Fourier superspace.

This leads us to the symmetry issue. There seems to be no way around
hyperbolic symmetry if one wishes to set up non--linear $\sigma$
models. Here, however, we were interested in exact, non--asymptotic
results. Although this requires the calculation of certain group
integrals, it simplifies the symmetry: We showed that compact
supergroups are sufficient to construct supersymmetric integral
representations of the correlation functions for arbitrary positions
of the imaginary increments. This is a more formal justification of a
procedure which we have been using in previous work. We conclude that
hyperbolic symmetry is a necessity for non--linear $\sigma$ models,
but not for supersymmetric theories in general.  Nevertheless, even
though mathematics can be nicer than one expects, it is an interesting
challenge to also derive those supersymmetric integral representations
from a version of the theory in terms of non--compact groups.

We gave two general results for the correlation functions. The first
one involves certain correlations to which we refer as fundamental,
while the second one is only in terms of eigenvalue integrals. Both
results are valid for arbitrary rotation invariant Random Matrix
Ensembles. In particular, no factorization property of the probability
density has to be assumed.

\section*{Acknowledgments}

I thank Gernot Akemann and Heiner Kohler for fruitful discussions.  I
acknowledge financial support from Det Svenska Vetenskapsr\aa det.

\appendix

\section{Equality of the Traces}
\label{AppA}

The assertion~(\ref{tra}) is obviously correct for $m=1$, because we have
\begin{eqnarray}
\tr K &=& \sum_{p=1}^k \tr \left( L_p z_pz_p^\dagger -  \zeta_p\zeta_p^\dagger \right) 
          =  \sum_{p=1}^k \left( L_p z_p^\dagger z_p +  \zeta_p^\dagger\zeta_p \right)
                                                        \nonumber\\
        &=& \trg B \ .
\label{a1}
\end{eqnarray}
For $m=2,3,\ldots$, we find
\begin{eqnarray}
\tr K^m &=& \tr ALA^\dagger \cdots ALA^\dagger 
                                                        \nonumber\\
             &=& \tr AL^{1/2}L^{1/2}A^\dagger AL^{1/2} \cdots 
                                 L^{1/2}A^\dagger AL^{1/2} L^{1/2}A^\dagger
                                                        \nonumber\\
             &=& \tr AL^{1/2} B^{m-1} L^{1/2}A^\dagger \ .
\label{a2}
\end{eqnarray}
Without anticommuting variables, we could now simply use the
invariance of the trace under cyclic permutation and would arrive at
the desired result~(\ref{tra}), but with an ordinary trace also on the
right hand side. To carefully account for the anticommuting variables,
we write $C=B^{m-1}$ and introduce the upper indices ${\rm (c1)}$ and
${\rm (c2)}$ in boson--fermion block notation for the commuting
variables as well as ${\rm (a12)}$ and ${\rm (a21)}$ for the
anticommuting ones.  We obtain
\begin{eqnarray}
\tr K^m &=& \sum_{p,q} \tr \left( 
                     z_p L_p^{1/2} C_{pq}^{\rm (c1)} L_q^{1/2} z_q^\dagger -
                     z_p L_p^{1/2} C_{pq}^{\rm (a12)} \zeta_q^\dagger \right.
                                                        \nonumber\\
           & & \qquad\qquad + \left. \zeta_p C_{pq}^{\rm (a21)} L_q^{1/2} z_q^\dagger -
                     \zeta_p C_{pq}^{\rm (c2)} \zeta_q^\dagger \right)
                                                        \nonumber\\
            &=& \sum_{p,q} \left( 
                     L_q^{1/2} z_q^\dagger z_p L_p^{1/2} C_{pq}^{\rm (c1)} +
                     \zeta_q^\dagger z_p L_p^{1/2} C_{pq}^{\rm (a12)} \right.
                                                        \nonumber\\
           & & \qquad\qquad + \left.
                     L_q^{1/2} z_q^\dagger \zeta_p C_{pq}^{\rm (a21)} +
                     \zeta_q^\dagger \zeta_p C_{pq}^{\rm (c2)} \right)
                                                        \nonumber\\
           &=& \sum_{p,q} \left( 
                     B_{pq}^{\rm (c1)} C_{pq}^{\rm (c1)} +
                     B_{pq}^{\rm (a12)} C_{pq}^{\rm (a21)} -
                     \left( B_{pq}^{\rm (a21)} C_{pq}^{\rm (a12)} +
                     B_{pq}^{\rm (c2)} C_{pq}^{\rm (c2)} \right) \right)
                                                        \nonumber\\
           &=& \trg BC = \trg BB^{m-1} = \trg B^m \ , 
\label{a3}
\end{eqnarray}
as claimed.

\section{Details of the Spectral Decomposition}
\label{AppB}

The matrix $K$ is ordinary Hermitean, although anticommuting variables
are present. In particular, all inverses of the matrix elements
$K_{n^\prime n}$ exist. The eigenvalues $Y_n$ are thus uniquely
defined. Moreover the diagonalizing matrix
$\widetilde{V}=[\widetilde{V}_1 \ \cdots \ \widetilde{V}_N]$
introduced in Eq.~(\ref{diagk}) is ordinary unitary and in ${\rm
SU}(N)$, and the corresponding eigenvectors $\widetilde{V}_n$ are
orthonormal and have commuting elements only. This might seem to be at
odds with the form~(\ref{eva}) of the eigenvectors $V_n$. In the
second representation of the supervectors $w_n$, the $w_{np1}$ are
anticommuting and the $w_{np2}$ are commuting, such that all elements
of the vector $V_n$ are anticommuting, despite the fact that $K$ is an
ordinary matrix. To clarify this, we use that the $\widetilde{V}_n$
form a complete set and expand
\begin{eqnarray}
V_n = \sum_{n^\prime=1}^N \gamma_{nn^\prime}\widetilde{V}_{n^\prime} \ ,
\label{b1}
\end{eqnarray}
where the coefficients $\gamma_{nn^\prime}$ are commuting in the first
and anticommuting in the second representation of the vectors $w_n$.
From the eigenvalue equation~(\ref{eva}), we find
$\widetilde{V}_{l}^\dagger KV_n = Y_n \widetilde{V}_{l}^\dagger V_n$.
Inserting the expansion~(\ref{b1}), we obtain
$(Y_l-Y_n)\gamma_{ln}=0$, which implies that the coefficients satisfy
$\gamma_{ln}=\gamma_n\delta_{ln}$ with new coefficients $\gamma_n$. Hence
we have
\begin{eqnarray}
V_n = \gamma_n \widetilde{V}_{n} \ ,
\label{b2}
\end{eqnarray}
such that the eigenvectors $V_n$ and $\widetilde{V}_n$ are
proportional to each other. The $V_n$ are orthogonal, but they cannot be
normalized in the standard way, if the $\gamma_n$ are
anticommuting. We emphasize that this causes no problem whatsoever.
The coefficients can be written as the scalar products
\begin{eqnarray}
\gamma_n =  \widetilde{V}_{n}^\dagger V_n = \widetilde{V}_{n}^\dagger AL^{1/2}w_n \ .
\label{b2a}
\end{eqnarray}
If $\gamma_n$ is anticommuting, it is has to be nilpotent, which means
that an integer $j$ exists such that $\gamma_n^j=0$. We notice that
$\gamma_n$ is not nilquadratic, i.e.~the number $j$ is here larger
than two, because $\gamma_n$ is a complicated linear combination of
nilquadratic anticommuting variables. In general, a linear combination
of $J$ nilquadratic anticommuting variables is nilpotent for every
$j>J+1$.  Moreover, we also deduce from Eq.~(\ref{b2})
\begin{eqnarray}
V_{n^\prime}^\dagger V_n = \gamma_n^*\gamma_n \delta_{n^\prime n}
\label{b2b}
\end{eqnarray}
as the orthogonality relation.

It is worthwhile to also collect more information about the supermatrix
$B$. According to the definition~(\ref{bdp}), it is non--Hermitean and
satisfies
\begin{eqnarray}
B^\dagger = L B L \ .
\label{b3}
\end{eqnarray}
One easily sees that 
\begin{eqnarray}
w_{p^\prime}^\dagger L w_p = \delta_{p^\prime p}
\label{b4}
\end{eqnarray}
is the corresponding orthonormality relation for the eigenvectors
$w_p$.  Being supervectors, these eigenvectors can always be properly
normalized to unity. The completeness relation reads
\begin{eqnarray}
\sum_{p=1}^{2k} w_p w_p^\dagger L = 1_{2k} \ ,
\label{b5}
\end{eqnarray}
where $1_{2k}$ is the $2k \times 2k$ unit matrix.

We construct a helpful alternative representation of the matrix $K$.
Employing the form~(\ref{eva}) and the completeness
relation~(\ref{b5}) we work out the expression
\begin{eqnarray}
\sum_{n=1}^{2k} V_n V_n^\dagger &=&
\sum_{n=1}^{2k} AL^{1/2} w_nw_n^\dagger (L^{1/2})^\dagger A^\dagger
            \nonumber\\
&=& \sum_{n=1}^{2k} AL^{1/2} w_nw_n^\dagger L L^{1/2} A^\dagger =
ALA^\dagger \ ,
\label{b6}
\end{eqnarray}
and by virtue of Eq.~(\ref{ksp}) we arrive at
\begin{eqnarray}
K = \sum_{n=1}^{2k} V_n V_n^\dagger \ . 
\label{b7}
\end{eqnarray}
This spectral decomposition is somewhat strange, because the
eigenvalues do not appear explicitly. However, useful results can be
deduced from it.  In the eigenvalue equation $KV_n=Y_nV_n$ it gives
together with the orthogonality relation~(\ref{b2b})
\begin{eqnarray}
Y_n = V_n^\dagger V_n = \gamma_n^* \gamma_n \ .
\label{b8}
\end{eqnarray}
Thus, the $k$ eigenvalues $Y_{p+k}=y_{p2}$ are products of two nilpotent
anticommuting variables.

Furthermore, one readily sees from the decomposition~(\ref{b7}) that
all eigenvalues $Y_n , \ n>2k$ which are different from $Y_p=y_{p1}$
and $Y_{p+k}=y_{p2}$ must be zero. As $K$ is Hermitean, one can
convince oneself in the usual way that the eigenvectors $V_n$ to
different eigenvalues are orthogonal. Let $V_n$ be an eigenvector to
an eigenvalue $Y_n$ with $n>2k$. We immediately conclude from
Eq.~(\ref{b7}) that $KV_n=0$ and hence $Y_n=0$.

Finally, we show in an alternative way that the $2k$ non--zero
eigenvalues $Y_n$ of $K$ coincide with the eigenvalues of $B$. We
write the eigenvalue equation as $Bw_n=b_nw_n$ and consider
the orthogonality relation
\begin{eqnarray} 
\delta_{n^\prime n} Y_n &=& V_{n^\prime}^\dagger V_n 
                                                                   \nonumber\\ 
        &=& w_{n^\prime}^\dagger (L^{1/2})^\dagger A^\dagger A L^{1/2} w_n 
        = w_{n^\prime}^\dagger (L^{1/2})^\dagger L^{-1/2} L^{1/2} A^\dagger A L^{1/2} w_n 
                                                                   \nonumber\\ 
        &=& w_{n^\prime}^\dagger L B  w_n 
        = b_n w_{n^\prime}^\dagger L w_n = b_n \delta_{n^\prime n} \ .
\label{ortho}
\end{eqnarray}
For $n=n^\prime$ we conclude $b_n=Y_n$ as claimed, and for $n\neq
n^\prime$ we observe that the orthogonalities of the eigenvectors
$V_n$ and $w_n$ mutually imply each other.

\section{A Supersymmetric Ingham--Siegel Integral}
\label{AppC}

For $2k\times 2k$ Hermitean supermatrices $\sigma$ and $\rho$, we wish
to calculate the integral $I(\rho)$, i.e.~the Fourier
transform~(\ref{foudetg}) of the superdeterminant. As $I(\rho)$ is
obviously an invariant function depending on eigenvalues only, we may
replace $\rho$ with $r$.  Up to a certain point, we can apply and
slightly extend the methods given in Ref~\cite{YF} for the case of
ordinary matrices.  Employing the notation of~\ref{AppA}, the matrix
$\sigma$ is viewed as consisting of the element $\sigma_{11}^{\rm
(c1)}$, the supervector $\vec{\sigma}_1=(\sigma_{21}^{\rm
(c1)},\ldots,\sigma_{k1}^{\rm (c1)},
\sigma_{11}^{\rm (a21)},\ldots,\sigma_{k1}^{\rm (a21)})$ with $k-1$
commuting and $k$ anticommuting variables, the complex conjugate
$\vec{\sigma}_1^\dagger$ and the $(2k-1)\times (2k-1)$ Hermitean
supermatrices $\widetilde{\sigma}$ containing all other matrix
elements. Because of
\begin{equation}
\detg\sigma^- = \detg\widetilde{\sigma}^- \, \left(\sigma_{11}^{{\rm (c1)}-} 
            + \vec{\sigma}_1^\dagger(\widetilde{\sigma}^-)^{-1}\vec{\sigma}_1\right)
\label{c2}
\end{equation}
the integral over $\sigma_{11}^{\rm (c1)}$ can easily be done with the
help of the residue theorem. Some care is needed, because the bilinear
form $\vec{\sigma}_1^\dagger(\widetilde{\sigma}^-)^{-1}\vec{\sigma}_1$
is an undetermined complex number due to the presence of the imaginary
increments. However, as the variables $\widetilde{\sigma}$ are only
parameters in the $\sigma_{11}^{\rm (c1)}$ integration, we may shift
the imaginary increments away, assuming that $\widetilde{\sigma}$ can
be inverted. The unitary supermatrix diagonalizing
$\widetilde{\sigma}$ can then be absorbed into the supervector
$\vec{\sigma}_1$. This makes the bilinear form
$\vec{\sigma}_1^\dagger(\widetilde{\sigma})^{-1}\vec{\sigma}_1$ real,
and the residue is well determined. The integral over the supervector
$\vec{\sigma}_1$ is then simply Gaussian and we find
\begin{eqnarray}
I(\rho) &\sim& \Theta(r_{11}) (ir_{11})^N \exp\left(-\varepsilon r_{11}\right) 
                                                                          \nonumber\\
      & & \qquad\qquad\qquad
               \int d[\widetilde{\sigma}] \, \exp\left(i\trg\widetilde{r}\widetilde{\sigma}\right) 
               \detg^{-(N-1)}\widetilde{\sigma}^- 
\label{c3}
\end{eqnarray}
with $\widetilde{r} = \diag(r_{21}, \ldots, r_{k1}, ir_{k2}, \ldots,
ir_{k2})$. We perform the calculation up to the normalization constant
which will be determined later on.  It should be noticed that the
presence of the anticommuting variables leads to some differences as
compared to the corresponding formula in Ref.~\cite{YF}. We can repeat
this step $k-1$ further times until all variables $\sigma_{pq}^{\rm
(c1)}$ and all anticommuting variables $\sigma_{pq}^{\rm (a21)}$ and
$\sigma_{pq}^{\rm (a12)}$ have been integrated out. This results in
\begin{eqnarray}
I(\rho) &\sim& \prod_{p=1}^k \Theta(r_{p1}) (ir_{p1})^N \exp\left(-\varepsilon r_{p1}\right) 
                                                      J(r_2) 
                        \nonumber\\  
J(r_2) &=& \int d[\sigma^{\rm (c2)}] \, \exp\left(i\tr r_2\sigma^{\rm (c2)}\right) 
                                                 {\det}^{N-k}\sigma^{\rm (c2)} \ .
\label{c4}
\end{eqnarray}
The remaining integral $J(r_2)$ is over the ordinary $k\times k$
Hermitean matrix $\sigma^{\rm (c2)}$. As no anticommuting variables
appear in the integrand, the inverse superdeterminant is identical to
the determinant in the numerator. This determinant does not contain
singularities anymore and thus we dropped the imaginary increments.
Upon introducing eigenvalue--angle coordinates for $\sigma^{\rm (c2)}$
and applying the Harish-Chandra--Itzykson--Zuber integral~(\ref{hciz})
for ${\rm U}(k)$, we are left with an integral over the eigenvalues
$s_p^{\rm (c2)}, \ p=1,\ldots,k$ given by
\begin{equation}
J(r_2) \sim \frac{1}{\Delta_k(r_2)} \int d[s^{\rm (c2)}] \, \Delta_k(s^{\rm (c2)}) 
               \exp\left(i\tr r_2s^{\rm (c2)}\right) {\det}^{N-k}s^{\rm (c2)} \ .
\label{c5}
\end{equation}
As the Vandermonde determinant appearing in the remaining integral
reads
\begin{equation}
\Delta_k(s^{\rm (c2)}) = \prod_{p<q}\left(s_p^{\rm (c2)}-s_q^{\rm (c2)}\right)
                                     = \det\left[\left(s_p^{\rm (c2)}\right)^{q-1}\right]_{p,q=1,\ldots,k}  \ ,
\label{c6}
\end{equation}
we can do the eigenvalue integrals and have
\begin{equation}
J(r_2) \sim \frac{1}{\Delta_k(r_2)} \, 
              \det\left[\frac{\partial^{N-k+q-1}\delta(r_{p2})}
                                  {\partial r_{p2}^{N-k+q-1}}\right]_{p,q=1,\ldots,k} \ .
\label{c7}
\end{equation}
The Vandermonde determinant $\Delta_k(r_2)$ cancels out. To see this
one may use the identity
\begin{equation}
\frac{\partial^{N-k+q-1}\delta(r_{p2})}
                                  {\partial r_{p2}^{N-k+q-1}} =
(-1)^{N-1} \frac{(N-k+q-1)!}{(N-1)!} 
              r_{p2}^{k-q}\frac{\partial^{N-1}\delta(r_{p2})}{\partial r_{p2}^{N-1}} \ ,
\label{c7a}
\end{equation}
which is easily derived by multiplying the right hand side with a well--behaved,
non--singular function and integrating by parts. Thus, we have
\begin{eqnarray}
\det\left[\frac{\partial^{N-k+q-1}\delta(r_{p2})}
                                  {\partial r_{p2}^{N-k+q-1}}\right]_{p,q=1,\ldots,k}
&\sim& \det\left[r_{p2}^{k-q}\frac{\partial^{N-1}\delta(r_{p2})}
                                  {\partial r_{p2}^{N-1}}\right]_{p,q=1,\ldots,k}
                                        \nonumber\\
&=& \det\left[r_{p2}^{k-q}\right]_{p,q=1,\ldots,k}
            \prod_{p=1}^k \frac{\partial^{N-1}\delta(r_{p2})}{\partial r_{p2}^{N-1}}
                                        \nonumber\\
&=& \Delta_k(r_2) 
            \prod_{p=1}^k \frac{\partial^{N-1}\delta(r_{p2})}{\partial r_{p2}^{N-1}} \ .
\label{c8}
\end{eqnarray}
Collecting everything we arrive at the first of the
results~(\ref{fdr}). The normalization constant $c_{Nk}$ is found by
integrating $I(\rho)$ together with a normalized Gaussian,
\begin{eqnarray}
& & C_{Nk}\left(\frac{(-1)^{N-1}(N-1)!}{i2\pi}\right)^k = I(0) 
                      = 2^{k(k-1)} \int d[\rho] \exp\left(-\trg\rho^2\right) I(\rho)
                                         \nonumber\\
& & \qquad\qquad = \int d[\sigma] \exp\left(-\frac{1}{4}\trg\sigma^2\right) \detg^{-N}\sigma^-
                   = 2^{-k(k-1)} \ .
\label{c9}
\end{eqnarray}
When calculating $I(0)$, we used $\Theta(0)=1/2$, as follows from a
careful inspection of the step leading from Eq.~(\ref{c2}) to
Eq.~(\ref{c3}). The second and the last equality sign in
Eq.~(\ref{c9}) are due to the Efetov--Wegner--Parisi--Sourlas
theorem~\cite{EFE83,PaSou,Weg,ConGro,ROT}, and the equality of the two
integrals is a direct consequence of the definition~(\ref{foudetg}).

\section{Properties of the Fundamental Correlations and the Fundamental Kernel}
\label{AppD}

The fundamental correlations~(\ref{corrfund}) are closely related, but
not identical to the superdeterminant, i.e.~to the Fourier
backtransform of the function $I(\rho)$, defined in
Eq.~(\ref{foudetg}). This is so, because the
Efetov--Wegner--Parisi--Sourlas
contributions~\cite{EFE83,PaSou,Weg,ConGro,ROT} for the function
$I(\rho)$ and for the whole generating function $Z_k(x+J)$ are
different. To gain further insights, we rewrite the fundamental
kernel~(\ref{kerfund}). We introduce two $2\times 2$ Hermitean
supermatrices $\overline{\sigma}$ and $\overline{\rho}$ with
eigenvalues $\overline{s}=\diag(s_{p1},is_{q2})$ and
$\overline{r}=\diag(r_1,ir_2)$, respectively. We then can cast
Eq.~(\ref{kerfund}) into the form
\begin{equation}
\widehat{C}^{\rm (fund)}(s_{p1},is_{q2}) =
-\frac{1}{\pi (s_{p1}^--is_{q2})} 
\left(\int d[\overline{\rho}]
        \exp\left(-i\trg\overline{\rho}\overline{\sigma}\right)  I(\overline{\rho}) - 1\right) \ ,
\label{d01}
\end{equation}
which is easily verified with the help of
formula~(\ref{hcizs}). Importantly, we have to subtract the
Efetov--Wegner--Parisi--Sourlas contribution, i.e.~unity in the
present case, from the integral. The integral over $\overline{\rho}$ is now
precisely the Fourier backtransform of Eq.~(\ref{foudetg}) and we have
\begin{eqnarray}
\widehat{C}^{\rm (fund)}(s_{p1},is_{q2}) &=&
-\frac{1}{\pi (s_{p1}^--is_{q2})} 
\left(\detg^{-N}\overline{\sigma}^- - 1\right) 
                \nonumber\\
&=& -\frac{1}{\pi (s_{p1}^--is_{q2})} \left(\left(\frac{is_{q2}}{s_{p1}^-}\right)^N - 1\right) 
                \nonumber\\
&=& \frac{1}{\pi (s_{p1}^-)^N} \frac{(s_{p1}^-)^N-(is_{q2})^N}{s_{p1}^--is_{q2}} \ ,
\label{d02}
\end{eqnarray}
which is the result~(\ref{kerfegs}). We notice that the fundamental
kernel and the fundamental correlation function coincide for
$k=1$. For $k>1$, this is not so and the calculation just presented
works for the fundamental kernel, but not for the whole fundamental
correlation function, because we do not know the corresponding
Efetov--Wegner--Parisi--Sourlas contributions explicitly.

To calculate the double integral~(\ref{kerfund}) directly, we apply
techniques developed in Refs.~\cite{TG,Guh93}. We integrate the $r_2$
integral $N-1$ times by parts and have
\begin{eqnarray}
\widehat{C}^{\rm (fund)}(s_{p1},is_{q2}) &=& - \frac{(-1)^{N-1}}{\pi (N-1)!}
\int\limits_{-\infty}^{+\infty} dr_1 \Theta(r_1) (ir_1)^N \exp\left(-ir_1s_{p1}^-\right) 
                          \nonumber\\
& & \qquad\qquad\qquad \frac{\partial^{N-1}}{\partial r_2^{N-1}} 
                 \frac{\exp\left(-ir_2s_{q2}\right)}{r_1-ir_2}\Bigg|_{r_2=0} \ .
\label{d1}
\end{eqnarray}
The $N-1$ fold derivative is computed with Leibnitz' rule according to
\begin{equation}
\frac{\partial^{N-1}}{\partial r_2^{N-1}} 
                 \frac{\exp\left(-ir_2s_{q2}\right)}{r_1-ir_2}\Bigg|_{r_2=0} 
= (-1)^N (N-1)! \sum_{n=0}^{N-1} \frac{(is_{q2})^n}{n! (ir_1)^{N-n}} \ ,
\label{d2}
\end{equation}
which yields Eq.~(\ref{kerfe}).

\section{Rederivation of the GUE Correlation Functions}
\label{AppE}

Again, we resort to techniques developed in Refs.~\cite{TG,Guh93}. For
a Gaussian probability density $P(H)\sim \exp(-\tr H^2)$, we obviously
have
\begin{equation}
P^{\rm (diag)}(H_{pp}) = \frac{1}{\sqrt{\pi}} \exp\left(-H_{pp}^2\right) 
\label{e1}
\end{equation}
in Eq.~(\ref{phfactor}). We plug this into Eq.~(\ref{corrafc}) and use
the first of Eqs.~(\ref{kerfe}). The $H_{(k+q)(k+q)} , \ q=1,\ldots,k$
integrals are then recognized as representations of the Hermite
polynomials
\begin{equation}
H_n(x)= \frac{2^n}{\sqrt{\pi}} 
               \int\limits_{-\infty}^{+\infty} \exp\left(-\xi^2\right) \left(z-i\xi\right)^n d\xi \ ,
\label{e2}
\end{equation}
while the $H_{pp} , \ p=1,\ldots,k$ integrals yield the generalized
Hermite functions
\begin{equation}
\widehat{H}_n(x)= \frac{(i2)^{n+1}}{\sqrt{\pi}} \exp\left(x^2\right) 
               \int\limits_0^\infty \exp\left(-\xi^2-i2x\xi\right) \xi^n d\xi \ .
\label{e3}
\end{equation}
These functions have already been introduced into the RMT context in
Ref.~\cite{Guh93}. They comprise both fundamental solutions of the
Hermite differential equation, the polynomials $H_n(x)$ and the
non--polynomial solutions $\widetilde{H}_n(x)$, in the form
\begin{equation}
\widehat{H}_n(x) = \widetilde{H}_n(x) + iH_n(x) \ ,
\label{e4}
\end{equation}
details can be found in Ref.~\cite{Guh93}. We also use the generalized
oscillator wave functions
\begin{equation}
\widehat{\varphi}_n(x) = \frac{1}{\sqrt{2^nn!\sqrt{\pi}}} 
                                         \exp\left(-\frac{x^2}{2}\right) \widehat{H}_n(x) 
                                       =  \widetilde{\varphi}_n(x) + i\varphi_n(x) \ .
\label{e5}
\end{equation}
Collecting everything, we find
\begin{eqnarray}
& & \int\limits_{-\infty}^{+\infty} dH_{pp} \frac{1}{\sqrt{\pi}} \exp\left(-H_{pp}^2\right) 
\int\limits_{-\infty}^{+\infty} dH_{(k+q)(k+q)} \frac{1}{\sqrt{\pi}} \exp\left(-H_{(k+q)(k+q)}^2\right) 
                   \nonumber\\
& & \qquad\qquad\qquad\qquad \widehat{C}^{\rm (fund)}(x_p-H_{pp},x_q-iH_{(k+q)(k+q)})
                   \nonumber\\
& & \qquad = \exp\left(\frac{x_q^2-x_p^2}{2}\right) 
                         \widehat{K}^{\rm (GUE)}(x_p,x_q) \ ,
\label{e6}
\end{eqnarray}
where
\begin{equation}
\widehat{K}^{\rm (GUE)}(x_p,x_q) = 
                                  \sum_{n=0}^{N-1}\widehat{\varphi}_n(x_p)\varphi_n(x_q)
\label{e7}
\end{equation}
is the GUE kernel, including the real part correlations. Hence, we arrive at
\begin{eqnarray}
\widehat{R}_k(x_1,\ldots,x_k) &=&
     \det\left[\exp\left(\frac{x_q^2-x_p^2}{2}\right)
                        \widehat{K}^{\rm (GUE)}(x_p,x_q)\right]_{p,q=1,\ldots,k}
                   \nonumber\\
      &=& \det\left[\widehat{K}^{\rm (GUE)}(x_p,x_q)\right]_{p,q=1,\ldots,k} 
                   \nonumber\\
      &=& \widehat{R}_k^{\rm (GUE)}(x_1,\ldots,x_k) \ ,
\label{e8}
\end{eqnarray}
which is the correct result.


\section*{References}

\begin{thebibliography}{30}

\bibitem{EFE83}
K.B. Efetov, Adv. Phys. 32 (1983) 53.

\bibitem{EFE97}
K.B. Efetov K,  {\it Supersymmetry in Disorder and Chaos},
Cambridge University Press, Cambridge  (1997).

\bibitem{VWZ}         
J.J.M. Verbaarschot, H.A. Weidenm\"uller and M. Zirnbauer,
Phys. Rep. {\bf 129} (1985) 367.

\bibitem{GMGW}    
T. Guhr, A. M\"uller--Groeling and H.A. Weidenm\"uller,  Phys. Rep. {\bf 299} (1998) 189.

\bibitem{Haake}   
F. Haake,  {\it Quantum Signatures of Chaos}, 2nd edition, (Springer Verlag, Berlin 2001).

\bibitem{TGenc}    
T. Guhr, {\it Random Matrix Theory in Physics},
Encyclopedia of Mathematical Physics, eds. J.-P. Françoise, G.L. Naber and
Tsou S.T, (Elsevier, Oxford 2006), vol.~4, pp.~338.

\bibitem{Mehta}   
M.L. Mehta, {\it Random Matrices}, 3rd edition, (Academic Press, New York 2004).

\bibitem{ZIR04}    
M. Zirnbauer, {\it Supersymmetry Methods in Random Matrix Theory},
Encyclopedia of Mathematical Physics, eds. J.-P. Françoise, G.L. Naber and
Tsou S.T, (Elsevier, Oxford 2006), vol.~5, pp.~151,
{\tt math-ph/0404057}

\bibitem{HaWe}
G. Hackenbroich and H.A. Weidenm\"uller, Phys. Rev. Lett. {\bf 74} (1995) 4418.

\bibitem{TGnorm}
T. Guhr, {\it Norm--dependent Random Matrix Ensembles in External Field and Supersymmmetry},
submitted for publication.

\bibitem{SW80}
L. Sch\"afer and F. Wegner, Z. Phys. {\bf B38} (1980) 113

\bibitem{TG}
T. Guhr, J. Math. Phys. {\bf 32} (1991) 336.

\bibitem{Guh96} 
T. Guhr, Phys. Rev. Lett. {\bf 76} (1996) 2258.

\bibitem{GUH4}
T. Guhr, Ann. Phys. (NY) {\bf 250} (1996) 145.

\bibitem{MK05}
K.A. Muttalib and J.R. Klauder, Phys. Rev. {\bf E71} (2005) 055101(R).

\bibitem{YF}
Y. Fyodorov, Nucl. Phys. {\bf B621} (2002) 643.

\bibitem{HC}
Harish-Chandra, Am. J. Math. {\bf 80} (1958) 241.

\bibitem{IZ}
C. Itzykson, J.B. Zuber, J. Math. Phys. {\bf 21} (1980) 411.

\bibitem{PPW}
T. Papenbrock, Z. Pluhar and H.A. Weidenm\"uller, {\tt cond-mat/0603525}.

\bibitem{GGT}
T. Guhr, Commun. Math. Phys. {\bf 176} (1996) 555.

\bibitem{PaSou}
G. Parisi and G. Sourlas, Phys. Rev. Lett. {\bf 43} (1979) 744.

\bibitem{Weg}
F. Wegner, Z. Phys. {\bf B49} (1983) 297.

\bibitem{ConGro}
F. Constantinescu and H.F. de Groote, J. Math. Phys. {\bf 30} (1989) 981.

\bibitem{ROT}         
M.J. Rothstein, Trans. Am. Math. Soc. {\bf 299} (1987) 387.

\bibitem{Guh93} 
T. Guhr, Nucl. Phys. {\bf A560} (1993) 223.

\bibitem{Bate}
H. Bateman,  {\it Higher Transcendental Functions}, Vol.~II,
A. Erd\'elyi (ed.), McGraw--Hill, New York, 1953.

\end{thebibliography}
\end{document}